%% file: tmi_author_accepted.tex
\documentclass[journal,twoside,web]{ieeecolor}
\usepackage{tmi}
\usepackage{cite}
\usepackage{amsmath,amssymb,amsfonts}
\usepackage{algorithmic}
\usepackage{graphicx}
\usepackage{textcomp}
\usepackage{algorithm}
\usepackage{array}
\usepackage{subcaption}
\usepackage{textcomp}
\usepackage{stfloats}
\usepackage{url}
\usepackage{verbatim}
\usepackage{tabularx, booktabs}
\usepackage{multirow}

\usepackage{float}

\newcommand{\imx}{\mathbf{x}}

\newcolumntype{Y}{>{\centering\arraybackslash}X}

\def\BibTeX{{\rm B\kern-.05em{\sc i\kern-.025em b}\kern-.08em
    T\kern-.1667em\lower.7ex\hbox{E}\kern-.125emX}}
\begin{document}
\begin{titlepage}
\centering
\Large
\vspace*{\fill}
{\bfseries Likelihood-Scheduled Score-Based Generative Modeling for Fully 3D PET Image Reconstruction\par}
\vspace{2\baselineskip}
This work has been accepted for publication in \emph{IEEE Transactions on Medical Imaging}.  
The author’s accepted manuscript is released under a CC-BY licence.  
For the Version of Record, see DOI: 10.1109/TMI.2025.3576483.
\vspace*{\fill}
\end{titlepage}
\title{Likelihood-Scheduled Score-Based Generative Modeling for Fully 3D PET Image Reconstruction}

\author{George Webber,~\IEEEmembership{Student Member,~IEEE}, Yuya Mizuno, Oliver D. Howes, Alexander Hammers, Andrew P. King, Andrew J. Reader
\thanks{ GW (e-mail: george.webber@kcl.ac.uk), APK and AJR are with the School of Biomedical Engineering and Imaging Sciences, King’s College London, UK. YM and ODH are with the Institute of Psychiatry, Psychology and Neuroscience, King’s College London, UK. ODH is also with the Institute of Clinical Sciences, Faculty of Medicine, Imperial College London. YM is also with South London and Maudsley NHS Foundation Trust, London, UK and the Department of Neuropsychiatry, Keio University School of Medicine, Tokyo, Japan. AH is with King’s College London and Guy's \& St Thomas' PET Centre.}
\thanks{ GW would like to acknowledge funding from the EPSRC Centre for Doctoral Training in Smart Medical Imaging [EP/S022104/1] and via a GSK studentship. This work was also supported in part by the Wellcome/EPSRC Centre for Medical Engineering [WT 203148/Z/16/Z], and in part by EPSRC grant number [EP/S032789/1].}
\thanks{ This study was also funded by Medical Research Council-UK (MC\_U120097115; MR/W005557/1 and MR/V013734/1), and Wellcome Trust (no. 094849/Z/10/Z) grants to Dr Howes and the National Institute for Health and Care Research (NIHR) Biomedical Research Centre at South London and Maudsley NHS Foundation Trust and King’s College London. The views expressed are those of the author(s) and not necessarily those of the NIHR or the Department of Health. }
\thanks{ For the purposes of open access, the authors have applied a Creative Commons Attribution (CC BY) licence to any Accepted Author Manuscript version arising, in accordance with King’s College London’s Rights Retention policy.}
}

\markboth{TMI - Author Accepted Manuscript}%
{Webber \textit{et al.}: Likelihood-Scheduled Score-Based Generative Modeling for Fully 3D PET Image Reconstruction}

\maketitle

\begin{abstract}
Medical image reconstruction with pre-trained score-based generative models (SGMs) has advantages over other existing state-of-the-art deep-learned reconstruction methods, including improved resilience to different scanner setups and advanced image distribution modeling. SGM-based reconstruction has recently been applied to simulated positron emission tomography (PET) datasets, showing improved contrast recovery for out-of-distribution lesions relative to the state-of-the-art. However, existing methods for SGM-based reconstruction from PET data suffer from slow reconstruction, burdensome hyperparameter tuning and slice inconsistency effects (in 3D). In this work, we propose a practical methodology for fully 3D reconstruction that accelerates reconstruction and reduces the number of critical hyperparameters by matching the likelihood of an SGM's reverse diffusion process to a current iterate of the maximum-likelihood expectation maximization algorithm. Using the example of low-count reconstruction from simulated $[^{18}$F]DPA-714 datasets, we show our methodology can match or improve on the NRMSE and SSIM of existing state-of-the-art SGM-based PET reconstruction while reducing reconstruction time and the need for hyperparameter tuning. We evaluate our methodology against state-of-the-art supervised and conventional reconstruction algorithms. 
Finally, we demonstrate a first-ever implementation of SGM-based reconstruction for real 3D PET data, specifically $[^{18}$F]DPA-714 data, where we integrate perpendicular pre-trained SGMs to eliminate slice inconsistency issues.
\end{abstract}


\section{Introduction}

Positron emission tomography (PET) is a nuclear medicine imaging technique used widely in clinical practice and research to image functional processes within the body \cite{bailey_positron_2005}. PET scans involve exposure to ionizing radiation from injecting a radioactive tracer, and this can be reduced by reducing the radioactive counts administered.
However, low-count data suffers from high levels of Poisson noise, leading to visually noisy images when conventional model-based reconstruction methods are used \cite{boellaard_standards_2009}. Deep learning methods have been proposed to compensate for the poor signal-to-noise ratio in measured low-count data \cite{reader_ai_2023, reader_deep_2021}. 

Most work in deep-learned PET reconstruction utilizes supervised deep learning, where a mapping is directly learned from low-dose PET data (e.g. sinograms) to high-quality images, either with \cite{mehranian_model-based_2020, guazzo_learned_2021} or without \cite{zhu_image_2018, haggstrom_deeppet_2019} advance knowledge of the fixed PET forward model.
 
A recent trend in medical image reconstruction is to leverage a score-based generative model (SGM) that has been pre-trained on a relevant image dataset as an unsupervised prior\cite{webber_diffusion_2024}. To perform unsupervised SGM-based reconstruction, the generative steps of the SGM are interleaved with reconstruction steps to encourage consistency between the generated image and the measured data \cite{chung_score-based_2022}. For 3D reconstruction, an SGM is typically pre-trained on diverse 2D transverse slices \cite{chung_solving_2023}, and the score-based prior is applied to these planes, while conventional regularization ensures slice consistency in the axial direction \cite{chung_solving_2023, singh_score-based_2024}.

Unlike supervised reconstruction methods, unsupervised SGM-based reconstruction only needs unpaired high-quality images for training, decoupling scanner-specific factors and improving generalizability \cite{chung_score-based_2022}. This simplifies training and allows greater flexibility at inference with varied dose levels and scanner parameters, though it may be less task-specific than supervised learning.

Some existing work has shown promise for SGM-based reconstruction of PET data \cite{singh_score-based_2024, hu_unsupervised_2024, webber_generative-model-based_2024}. However, existing SGM-based reconstruction methods suffer from issues including long reconstruction times and the need for burdensome hyperparameter tuning\cite{webber_diffusion_2024, hou_fast_2024}. In 3D, methods also suffer from inconsistency or blurring between axial slices (due to only applying the score-based prior in the transverse planes) \cite{singh_score-based_2024}.

\begin{figure*}[t]
    \centering
    \includegraphics[width=\textwidth]{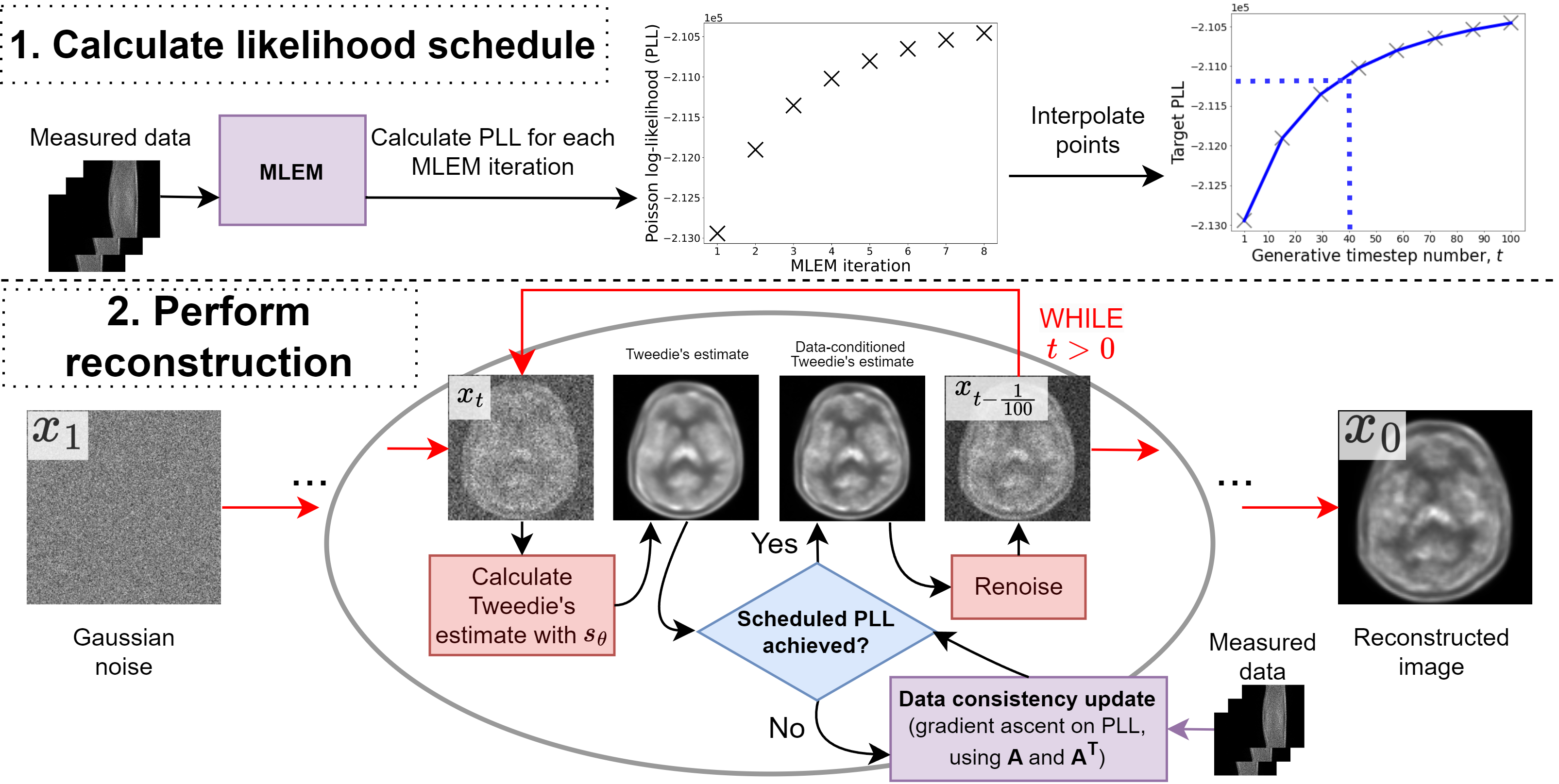}
    \caption{Our proposed likelihood-scheduling methodology for SGM-based PET image reconstruction.}
    \label{fig:methods_explainer}
\end{figure*}

In this work, we propose a likelihood-scheduling mechanism for SGM-based reconstruction to address the aforementioned issues of burdensome hyperparameter tuning, slice inconsistency and slow reconstruction. Our method first runs the maximum likelihood expectation maximization algorithm (MLEM) to generate a ``likelihood schedule" for a given set of measured sinogram data (see Fig. \ref{fig:methods_explainer}). The likelihood schedule is then integrated into the reverse diffusion process of an SGM-based reconstruction, enabling dynamic adjustment of the balance between the prior and the likelihood contributions. By integrating this development with SGMs trained on perpendicular slice orientations \cite{lee_improving_2023}, our method eliminates the slice inconsistency issue while reducing the number of critical regularization hyperparameters from 4 to 1.

Previous methods have implicitly altered the balance between likelihood and prior via regularization hyperparameters; our proposal is the first to investigate choosing the target likelihood upfront, providing samples from the posterior distribution of image reconstruction conditioned on both a likelihood value and noisy measured data.

We conduct numerical experiments to validate our method's effectiveness on low-count PET data, using the example of simulated 2D $[^{18}$F]DPA-714 radiotracer distributions, and evaluate its performance against state-of-the-art conventional, supervised, and unsupervised SGM-based reconstruction algorithms. We then extend to the 3D case, showing quantitative and qualitative results for real fully 3D PET reconstruction.

This work makes the following contributions:
\begin{itemize}
    \item We propose a principled and efficient mechanism for dynamically balancing SGM denoising steps with likelihood update steps for image reconstruction. Our methodology enables direct sampling of possible reconstructions at a fixed likelihood value.
    \item We show our method has a significantly lower hyperparameter selection burden than the state-of-the-art for unsupervised SGM-based reconstruction without compromising reconstruction accuracy.
    \item We resolve slice inconsistency issues in 3D by integrating our method with perpendicular pre-trained 2D SGMs and demonstrate the first-ever fully 3D PET reconstructions from real data using SGM-based reconstruction, specifically from low-count data acquired with radiotracer $[^{18}$F]DPA714.
\end{itemize}

\section{Background}

\newcommand{\imq}{\mathbf{q}}
\newcommand{\imA}{\mathbf{A}}
\newcommand{\imy}{\mathbf{y}}
\newcommand{\imb}{\mathbf{b}}
\newcommand{\imm}{\mathbf{m}}
\newcommand{\score}{\nabla_{\imx} \log p_t(\imx_t) }
\newcommand{\scorecond}{\nabla_{\imx} \log p_t(\imx_t | \imm) }
\newcommand{\scorelike}{\nabla_{\imx} \log p_t(\imm | \imx_t) }
\newcommand{\scorenn}{ \mathbf{s}_{\mathbf{\theta}}(\imx_t, t)}
\newcommand{\stdnorm}{ \mathcal{N}(\mathbf{0}, \mathbf{I}) }
\newcommand{\tweedie}{\hat{\imx}_0(\imx_t)}
\newcommand{\nmlem}{N_{\text{MLEM}}}
\newcommand{\ngen}{N_{\text{gen}}}
\newcommand{\nsubsets}{N_{\text{subsets}}}

\subsection{PET reconstruction} \label{sec:pet_recon}

Reconstructing an image from PET emission data is an inverse problem \cite{reader_deep_2021}. The true mean $\imq$ of noisy measurements $\imm$ (e.g. a sinogram) may be modeled as 
\begin{equation}
    \imq = \imA\imx + \imb \;,
\end{equation}
where $\imx$ represents the true radiotracer distribution, $ \imA $ represents our system model and $ \imb $ models scatter and randoms components. The system model $\imA$ includes the image space point spread function (PSF), projection between image and sinogram space as well as attenuation and normalization modeling.

PET emission data is generated as a set of random discrete emissions from radionuclides, and therefore follows a Poisson noise model. MLEM \cite{shepp_maximum_1982} is a convergent iterative algorithm that maximizes the Poisson log-likelihood (PLL) of emission data with respect to an image estimate, given by 
\begin{equation}
    L(\imx | \imm) = \sum_{i=1}^{k} m_i \log([\imA\imx + \imb]_i) - [\imA\imx + \imb]_i - \log(m_i!)\; .
\end{equation}
However, with low-count data and a high-dimensional $\imx$, the maximum likelihood estimate overfits to noisy measurement data. It is standard to compensate for this reduction in signal by conditioning on an image-based prior, thereby regularizing the reconstruction, via algorithms such as maximum \textit{a posteriori}-expectation maximization (MAP-EM) \cite{levitan_maximum_1987}. For this purpose, let $ p(\imx) $ be the prior probability density for an image $ \imx $.

Such algorithms may be accelerated by partitioning sinograms into subsets and seeking the maxima of a set of corresponding sub-objectives, e.g. leading to Ordered-Subset Expectation Maximization (OSEM) \cite{hudson_accelerated_1994} and Block-Sequential Regularization Expectation Maximization (BSREM) \cite{de_pierro_fast_2001} for MLEM and MAP-EM respectively.

\subsection{Score-based generative models (SGMs)}

Score-based generative modeling is a generative deep learning framework that enables state-of-the-art modeling and sampling from the probability distribution $ p(\imx) = p_0(\imx) = \pi$ of a set of images \cite{song_generative_2019, sohl-dickstein_deep_2015, ho_denoising_2020, song_score-based_2020}. SGMs work by reversing a diffusion stochastic differential equation (SDE) that maps the initial distribution $ p(\imx) $ to a known distribution. In this paper, we consider the variance-preserving It\^{o} SDE that maps to the known distribution of Gaussian noise \cite{ho_denoising_2020}
\begin{equation}
    d\imx_t = -\frac{1}{2}\beta(t)\imx_t dt + \sqrt{\beta(t)}d\mathbf{w}_t
\end{equation}
where $\{\imx_t\}_{t \in [0,1]}$ is a stochastic process indexed by time $t$ and $ \{\mathbf{w}_t\}_{t \in [0,1]}$ is a standard Wiener process (multivariate Brownian motion). For each $t$, $\imx_t$ has associated density $p_t(\imx_t)$. The function $ \beta : [0,1] \rightarrow \mathbb{R}$ is chosen such that $ p_1 \approx \stdnorm$ (in this paper we fix $ \beta(t) = 0.1 + 19.9t$).

Anderson \cite{anderson_reverse-time_1982} gives the corresponding reverse-time SDE as 
\begin{equation}
    d\imx_t = \left[ -\frac{1}{2}\beta(t) - \beta(t) \score \right] dt + \sqrt{\beta(t)} d\mathbf{\bar{w}}_t
\end{equation}
where $\{\mathbf{\bar{w}}_t\}_{t \ge 0}$ is the time-reversed Wiener process and the term $ \nabla_{\imx} \log p_t(\imx_t)$ is the score function. To computationally simulate the reverse SGM, we train a noise-level-dependent neural network $\scorenn$, parameterized by  $\mathbf{\theta}$, to approximate the score function. This is achieved with Denoising Score Matching (DSM) \cite{vincent_connection_2011}, yielding the optimization problem
\begin{multline}\label{eq:DSM}
    \min_{\mathbf{\theta}} \big\{ \mathbb{E}_{t \sim U[0,1]} \mathbb{E}_{\imx_0 \sim \pi} \mathbb{E}_{\imx_t \sim p_t(\imx_t | \imx_0) } \bigl[  \\ \| \scorenn - \nabla_{\imx} \log p_t(\imx_t|\imx_0) \|_2^2 \; \bigl]     \big\} \;.
\end{multline}

Sampling from this generative model begins with sampling $\imx_1 \sim \stdnorm$. We then use the learned score model $\scorenn$ as a surrogate for the score function $ \score$, and simulate the resulting reverse SDE backwards in time (with a numerical solution such as Euler-Marayama schemes or predictor-corrector methods \cite{song_score-based_2020}), starting from $\imx_1$. 

DDIMs (Denoising Diffusion Implicit Models) \cite{song_score-based_2020} were introduced to allow faster sampling by reducing the necessity to simulate the SDE with a fine time-grid to produce high-quality samples. DDIMs utilize Tweedie's estimate \cite{efron_tweedies_2011} of the expectation $\mathbb{E}[\imx_0|\imx_t]$ using the score model as 
\begin{equation}
\mathbb{E}[\imx_0|\imx_t] = \frac{\imx_t + \nu_t^2 \score}{\gamma_t} \approx \frac{\imx_t + \nu_t^2\scorenn }{\gamma_t} := \tweedie
\end{equation}
where positive scalars $\gamma_t$ and $\nu_t^2$ are coefficients that may be derived from $\beta$ (see Singh \textit{et al.} \cite{singh_score-based_2024} for details). DDIM uses Tweedie's estimate and the current iterate $\imx_t$ to accelerate sampling with a non-Markovian sampling update rule
\begin{equation}\label{eq:ddim_sampling}
    \imx_{t_{k-1}} \sim \gamma_{t_{k-1}} \hat{\imx}_0(\imx_{t_k}) - \nu_{t_k}\sqrt{\nu^2_{t_{k-1}} - \eta^2_{t_k}} \scorenn + \eta_{t_k} \stdnorm 
\end{equation}
where $\eta_t$ is stochasticity (fixed at 0.1 for this work).

\subsection{PET reconstruction with SGMs}\label{sec:background_pet_with_sgm}

To solve the PET reconstruction problem with an SGM, we simulate the reverse diffusion process with an approximation of the conditional score $\scorecond$, allowing us to sample from $p(\imx | \imm)$. To approximate $\scorecond$, we decompose into prior and likelihood terms by Bayes' law as
\begin{equation}
    \begin{split}
        \scorecond &= \score + \scorelike\\
        &\approx \scorenn + \scorelike
    \end{split}
\end{equation}
and approximate the second term $ \scorelike $. 

While direct approximations to $ \scorelike $ have been investigated \cite{chung_decomposed_2023, singh_score-based_2024}, Singh \textit{et al.} find these too inefficient or inaccurate for 3D PET reconstruction \cite{singh_score-based_2024}. Several works \cite{zhu_denoising_2023, chung_decomposed_2023} have instead modified the DDIM sampling rule (\ref{eq:ddim_sampling}) for conditional generation, implicitly approximating the conditional likelihood by enforcing data consistency on Tweedie's estimate. These approaches calculate Tweedie's estimate $ \hat{\imx}_0(\imx_t) $ of the fully-denoised sample $\imx_0$, update $ \hat{\imx}_0(\imx_t) $ with an iterative data consistency scheme, and then add back Gaussian noise according to the DDIM update rule (\ref{eq:ddim_sampling}).

\section{Related work}

Following seminal works by Chung \textit{et al.} \cite{chung_score-based_2022, chung_solving_2023}, the integration of an image prior learned by SGMs with image reconstruction has proved effective in different medical imaging modalities - for a review see Webber \& Reader \cite{webber_diffusion_2024}.

Singh \textit{et al.} \cite{singh_score-based_2024} were the first to show simulated results for PET reconstruction using SGMs, demonstrating improved metrics of image quality and the ability to better recover simulated lesions from 3D phantom fluorodeoxyglucose (FDG) PET scans. Xie \textit{et al.} \cite{xie_joint_2024} considered joint reconstruction of PET-MR data utilizing a dual-domain diffusion process to show improvements over supervised learning methods on 2D sinograms simulated from real FDG PET images.

Recently, Hu \textit{et al.} \cite{hu_unsupervised_2024} showed results for unsupervised SGM-based reconstruction with simulated ultra-low dose FDG PET, outperforming conventional model-based iterative reconstruction (MBIR) methods.

At present, the only method shown in 3D for PET is Singh \textit{et al.}'s \cite{singh_score-based_2024} approach applying a pre-trained SGM to parallel axial slices, and a relative difference prior (RDP) to encourage consistency in the transverse direction. A general method to decompose 3D SGM-based reconstruction into multiple perpendicular 2D reconstruction problems has been proposed by Lee \textit{et al.} \cite{lee_improving_2023}.

\subsection{Motivation}

Singh \textit{et al.}\cite{singh_score-based_2024} propose PET-DDS, an adaptation of Decomposed Diffusion Sampling (DDS)\cite{chung_decomposed_2023} to the case of non-negative PET images with high dynamic range.

PET-DDS uses a modified DDIM sampling rule (see Section \ref{sec:background_pet_with_sgm}),  enforcing data consistency on Tweedie's estimate via gradient ascent steps on a MAP proximal objective. This objective has three terms: the PLL $L_j$ for subset $j$, an anchor term to prevent straying too far from the diffusion output and an axial relative difference prior (RDP) for 3D reconstruction.

When implementing PET-DDS, we empirically found that when reconstructing from ${\sim}10\times$ fewer counts than Singh \textit{et al.}, our log-likelihood gradients were large enough to prevent convergence of the proximal update. Therefore, we found it necessary to introduce the hyperparameter $ \delta $ to relax the rate of gradient ascent towards the reconstruction objective.

PET-DDS is a principled methodology that delivers high-quality reconstructions, but it has a number of shortcomings. PET-DDS has many hyperparameters to optimize, including: strength of MAP regularization $\lambda_{\text{MAP}}$; number of MAP iterations per generative step $N_{\text{MAP}}$; gradient ascent step size $\delta$; strength of RDP regularization $\lambda_{\text{RDP}}$; and, number of BSREM subsets $\nsubsets$. The first three of these depend on the time discretization and number of counts in the measured data.
Ideally, consistent hyperparameters across time discretizations would simplify hyperparameter tuning and support performing either fine- or coarse-grained reconstructions. 

Furthermore, when used in practice, convergence to PET-DDS's proximal objective is not achieved, and so the hyperparameters $\lambda_{\text{MAP}}$, $N_{\text{MAP}}$ and $\delta$ primarily act as proxies for the balance between the likelihood and the prior.

Additionally, using a constant likelihood strength across generative steps may be computationally inefficient. It is clear that likelihood steps early in the reverse diffusion process are less impactful than those later in the process, as the random noise added has more of an information-removal effect.
This motivates varying the strength of likelihood update at different generative steps for efficiency or quality improvements.

Lastly, existing methods for 3D reconstruction such as PET-DDS utilize a pre-trained SGM applied to axial slices through the reconstruction volume. This necessitates the inclusion of the transverse RDP, which has a smoothing effect on the reconstruction that causes an undesirable loss of detail.

\section{Proposed approach}

\subsection{Problem formulation}\label{sec:formulation}

Suppose $D$ is the probability distribution over images learned by a pre-trained SGM. Let $c$ be a real scalar. Then, we seek to solve the following problem:
\begin{equation}
    \text{`` Sample } \imx \text{ such that } \imx \sim D \text{ and } L(\imx | \imm) = c \text{ "}\;. 
\end{equation}

\begin{figure}
    \centering
    \includegraphics[width=\linewidth]{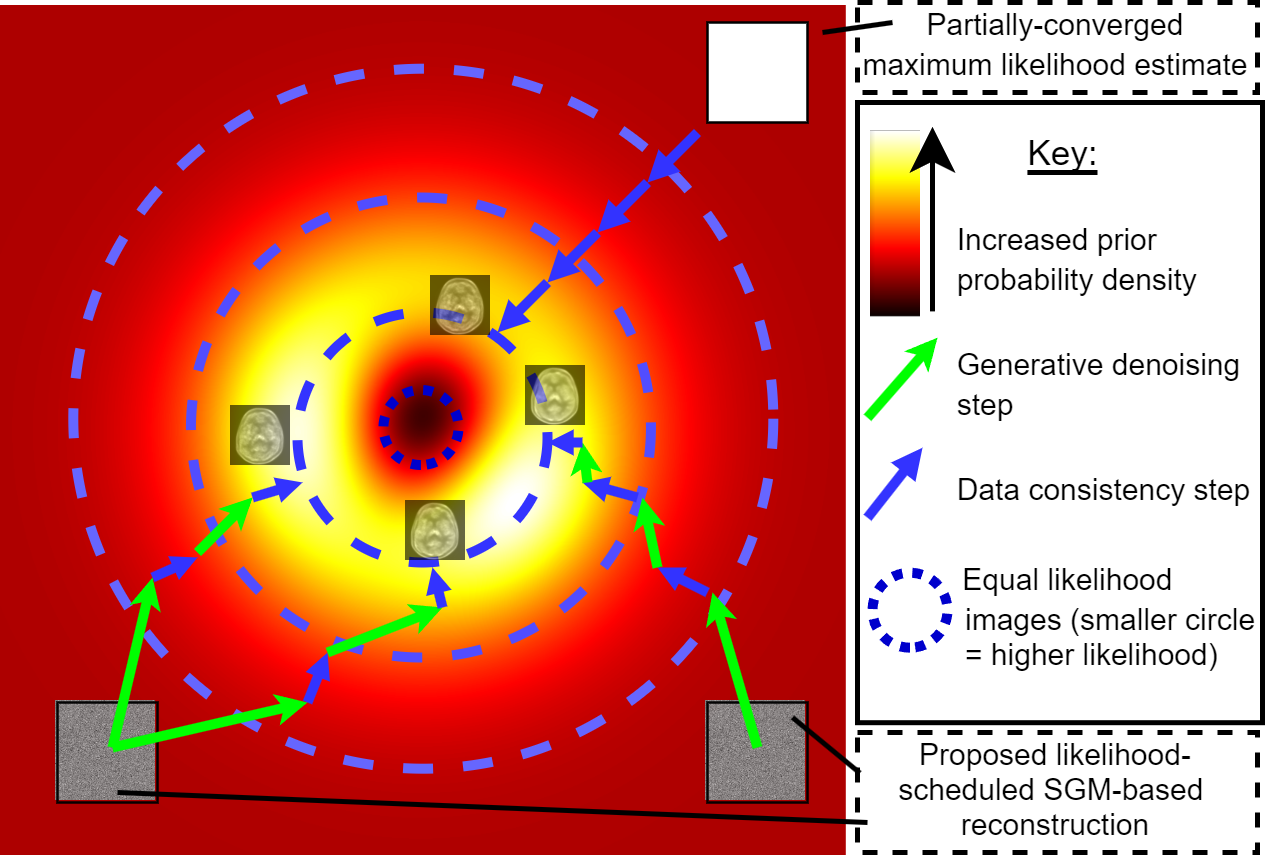}
    \caption{Explanatory figure showing the prior and likelihood values of reconstruction iterates for MLEM versus our method. The degeneracy of PLL means that there exists a manifold of equal likelihood images that we may sample according to the prior probability density learned by the SGM.}
    \label{fig:iso-manifold}
\end{figure}

This problem formulation may be viewed as sampling from a manifold of fixed likelihood images (see Fig. \ref{fig:iso-manifold}), as weighted by their probability under the learned prior distribution.

For this problem to be meaningful, $c$ should be chosen such that clinically-relevant images exist with log-likelihood equal to $c$. In this work, we choose $c$ by computing a clinically-relevant image $ \imx_{\text{MLEM}} $ with the MLEM algorithm, and setting $c = L( \imx_{\text{MLEM}} | \imm)$. This is a desirable selection, as solving the above problem would lead to sampled images that are equally consistent with measured sinograms as MLEM images, but without the issues of early-terminated MLEM (chiefly a lack of resolved detail).

\subsection{Likelihood-scheduling for SGM-based reconstruction}

We propose to solve the problem defined in Section \ref{sec:formulation} with a dynamic data consistency update that matches the likelihood of reconstruction iterates at each reverse diffusion step to a pre-computed `likelihood schedule'. For a visual explanation, see Fig. \ref{fig:methods_explainer}.

Firstly, we perform an MLEM reconstruction from sinogram data and record the PLL values for each image iterate (using $\imA$). We then linearly interpolate the PLL values into an $\ngen$-valued `likelihood schedule', where $\ngen$ is the number of generative denoising steps used in an SGM-based reconstruction.

Then, for the $i^{th}$ generative step, we first perform a single Tweedie denoising step, which we normalize with Singh \textit{et al.}'s measurement-normalization procedure \cite{singh_score-based_2024}. Next, we perform gradient ascent on Tweedie's estimate until the log-likelihood of our estimate exceeds the $i^{th}$ value in our likelihood schedule. This process occurs in pixel-space, utilizing $\imA$ and $\imA^{T}$. We then reapply noise to return a noisy iterate in accordance with the DDIM sampling rule (\ref{eq:ddim_sampling}).

The number of gradient ascent steps used in our algorithm is dynamic. The final reconstruction will have a PLL value within one gradient ascent step of a conventional MLEM reconstruction's PLL value.

This method has just two critical hyperparameters to tune: 1) $\nmlem$ the number of MLEM iterations used to determine the maximum end-point PLL value and, 2) the SGM's time discretization, i.e. $\ngen$ the number of generative steps used in the reverse diffusion process. Crucially, $\nmlem$ controls the relative balance of the prior and the likelihood (with larger $\nmlem$ giving greater emphasis to the likelihood), while $\ngen$ independently controls the number of diffusion timesteps used (and thereby trades off reconstruction speed against accuracy).

Furthermore, it is easier to tune the critical hyperparameter $\nmlem$, as it may be set to the number of EM updates for standard clinical reconstructions. All unsupervised PET reconstruction approaches involve heuristic regularization hyperparameters; a particular strength of our approach is to align this heuristic with an existing clinically-accepted and vendor-recommended heuristic, i.e. $\nmlem$.

The proposed approach is also more flexible to differing numbers of generative timesteps, as a likelihood schedule may be produced and adhered to for any number of generative timesteps. While hyperparameters such as the gradient ascent step size $\delta$ may still be specified, within reasonable bounds $\delta$ only controls the accuracy with which the likelihood schedule is conformed to, and not the balance of likelihood and prior. This is fundamentally different from gradient-ascent scheduling approaches such as linear annealing, which do not reduce the number of hyperparameters required.

\subsection{Adaptations to 3D}

To adapt our method to 3D, one could incorporate the axial-only RDP utilized by PET-DDS by replacing the likelihood schedule with an analogous ``objective schedule" consisting of the sum of the likelihood term and the RDP term. However, as discussed, the axial RDP causes undesirable blurring, particularly for sagittal and coronal slices.

We instead take inspiration from the approach of Lee \textit{et al.} \cite{lee_improving_2023}, by leveraging pre-trained SGMs trained on slices from orthogonal orientations. Namely, we pre-train three SGMs $ \mathbf{s}_{\mathbf{\theta}}^{\text{(sag)}}, \mathbf{s}_{\mathbf{\theta}}^{\text{(cor)}}, \mathbf{s}_{\mathbf{\theta}}^{\text{(tra)}} $ on diverse high-quality slices in the sagittal, coronal and transverse orientations respectively. During reconstruction, we apply each SGM to slices in its respective orientation and calculate the score $ \scorenn $ as the average of the score vectors output by the pre-trained SGMs. Note that this methodology differs from Lee \textit{et al.} in that for each generative timestep, the sum of scores from 3 perpendicular SGMs is used rather than alternating the choice of score between 2 perpendicular SGMs. Alternating between directions leads to Tweedie's estimate iterates with slice inconsistency effects, which are eliminates gradually as many diffusion steps are taken. Our different approach was hence motivated by the desire to reduce the number of diffusion timesteps (due to the expense of the likelihood updates) while eliminating slice inconsistency effects throughout the diffusion process.

To accelerate our method in 3D, we take larger gradient ascent steps. However, this can potentially cause less accurate matching to the likelihood schedule. To counter this imprecision, where an image iterate's PLL overshoots the target PLL, we linearly interpolate between the penultimate and current iterates (using the penultimate and current PLL) to yield a final iterate that better matches the target PLL.

\section{Experimental setup}

\subsection{Baseline methods}

We implemented PET-DDS and our proposed methodology with the same forward model as three baseline methods:

\subsubsection{OSEM}

As discussed in Section \ref{sec:pet_recon}, OSEM \cite{hudson_accelerated_1994} is a model-based iterative method widely used on clinical scanners. In OSEM, expectation-maximization steps are taken with respect to subsets of the measured data, resulting in an accelerated version of MLEM. Regularization is implicitly achieved by early stopping before full convergence to the noisy maximum likelihood image estimate.

\subsubsection{MAP-EM}

A more sophisticated conventional iterative method is MAP-EM \cite{levitan_maximum_1987}, an iterative algorithm for maximizing a regularized Poisson likelihood function. For our implementation, we follow Wang \& Qi's formulation of patch-based edge-preserving regularization \cite{wang_penalized_2012}. In this formulation, at each iteration an image estimate $\imx^{OSEM}_{t+1}$ is computed via an OSEM update and a regularization image $ \imx^{reg}_{t+1} $ is calculated with Wang \& Qi's regularization function. The OSEM estimate and regularization image are then combined using the De Pierro update\cite{de_pierro_relation_1993} weighted by a scalar hyperparameter $\beta$.

\subsubsection{FBSEM-net: unrolled iterative deep-learned method}

FBSEM-net (deep learning PET reconstruction with forward-backward splitting expectation-maximization) \cite{mehranian_model-based_2020} is a state-of-the-art supervised deep learning PET reconstruction algorithm, that offers a principled approach to incorporating deep learning into physics-based reconstruction. This method unrolls the iterative MAP-EM algorithm, replacing the hand-crafted prior with a neural network that is learned from data. 

Following Mehranian \& Reader \cite{mehranian_model-based_2020}, for computational and memory efficiency we perform 30 burn-in iterations of OSEM with 4 subsets, followed by 12 FBSEM-net steps that simultaneously regularize and accelerate the reconstruction. The reconstruction target for training purposes is the ground truth.

We compare to two implementations with different neural architectures for the regularizing neural network: FBSEM-net, with a convolutional neural network (CNN) comparable to Mehranian \& Reader \cite{mehranian_model-based_2020}, and FBSEM-net-adv, with the same network architecture used for score-based learning (with a constant timestep input). Section \ref{sec:training_details} contains training details.

FBSEM-net-adv is included in this study as a strong baseline representing the performance of deep learning approaches on within-distribution datasets when sinogram training data is available. Despite its strong performance in 2D, FBSEM-net-adv is omitted in 3D, due to computational infeasibility to train on available hardware.

\subsection{PET forward operator}

Each reconstruction method made use of the same ParallelProj projector
\cite{schramm_parallelprojopen-source_2024}. The geometry of the scanner was modeled using the publicly available specifications of Siemens' Biograph mMR scanner. 

The provided normalization was for data that has been axially compressed to span 11 (with 5 or 6 central lines of response summed along the axial direction). To cope with this constraint, axial compression was explicitly modeled in the forward operator (shown to have a negligible effect on reconstruction quality by Belzunce \& Reader\cite{belzunce_assessment_2017}). In 2D, a 4.5mm full-width half-maximum (FWHM) Gaussian PSF was also used.

The full forward model used was therefore:
\begin{equation} \imA (\mathbf{x}) = N_{11}L_{11}C_{11}X_1P \imx
\end{equation}
where $\imx$ is an image estimate, $N_{11}$ is a span 11 sinogram of normalization factors, $L_{11}$ is a span 11 sinogram of attenuation factors, $C_{11}$ is a compression operator that converts a span 1 sinogram into span 11, $X_1$ is the ParallelProj projector and $P$ is the Gaussian PSF.

\subsection{3D real $[^{18}$F]DPA-714 data}\label{sec:clinical_data}

69 static $[^{18}$F]DPA-714 brain datasets (from the Inflammatory Reaction in Schizophrenia team at King's College London \cite{muratib_dissection_2021}) were used in this work. The radiotracer $[^{18}$F]DPA-714 is a second-generation translocator protein (TSPO) PET probe, which is used to perform brain-wide quantitative analysis of TSPO. The datasets used in this work were acquired from healthy control subjects.

The data had been previously acquired from 1-hour scans with the Siemens Biograph mMR, with approximately 200 MBq administered, with total counts in the range $2.9 \times 10^8 $ to $ 1.3 \times 10^9$. At full-count, high-quality images (voxel size 2.08626 mm $\times$ 2.08626 mm  $\times$ 2.03125 mm ; 3D image size $128 \times 128 \times 120$) were reconstructed with the scanner defaults (OSEM with 21 subsets, 2 iterations and no PSF).

For each dataset, Siemens' scanner-specific algorithms were used to produce normalization sinograms, compute attenuation maps from previously acquired paired CT scans, and approximate the distribution of scatter events.

\subsection{2D simulated data}\label{sec:sim_data}

In order to have knowledge of the ground truth, simulated data were used. 2D transverse slices of high-quality $[^{18}$F]DPA-714 PET images were used as ground truths, with each image obtained via full-count reconstruction with Siemens' implementation of the OSEM algorithm (with 21 subsets and 42 iterations, i.e. 2 full passes through the data). Forward projected data were obtained using a single axial ring, after rescaling each ground truth image slice such that the total count of simulated prompts matched the original estimate of true events. Corrective factors were modeled as purely attenuation and contamination was modeled as a constant background of 30\% of simulated prompts.

In 2D, the dose level was set to 60\% of a single direct plane sinogram, resulting in an average $3.14\times10^5$ counts per reconstructed slice. Poisson noise was then applied to the clean forward projected data.

\subsection{Training and validation for SGMs and FBSEM-net}\label{sec:training_details}
For each of the transverse, coronal and sagittal orientations, an SGM was trained with all non-empty 2D slices from 55 3D training datasets (clinical images). Each SGM was trained to minimize the DSM objective (\ref{eq:DSM}) for 100 epochs, a value identified by 5-fold cross-validation on the transverse training datasets. Training-time data augmentations such as rotation and translation were performed. The SGM architecture used was identical to Singh \textit{et al.} \cite{singh_score-based_2024}.

2D FBSEM-net instances were trained using transverse slices from the same 55 training datasets as the SGMs, with slices from one 3D validation dataset used for early stopping on the validation loss. This same dataset was used for validation of the reconstruction process with the pre-trained SGMs, as well as bias-variance assessments in 2D. An additional 13 datasets were reserved as test data. All deep learning methods were trained with Adam with learning rate $1\times 10^{-4}$.

\begin{figure*}[th!]
    \centering
    \includegraphics[width=\textwidth]{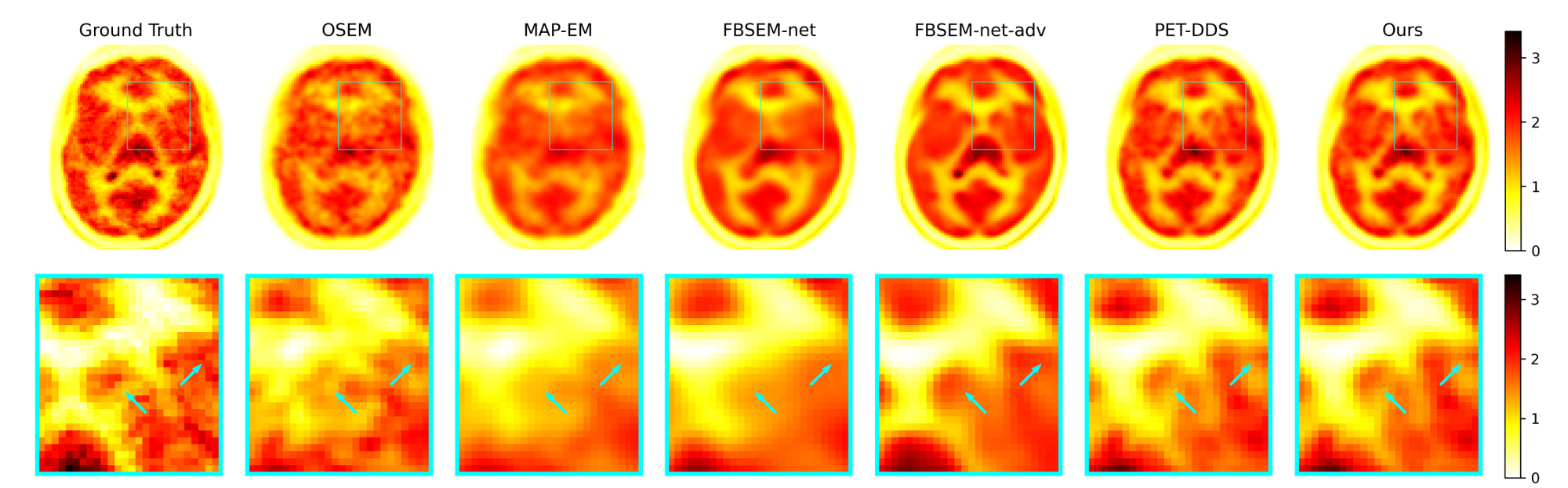}
    \caption{Reconstructions with each method from 2D simulated $[^{18}$F]DPA-714 data, with hyperparameters for each method chosen to minimize NRMSE on a validation dataset. Arrows point to key differences between reconstructions, including well-reconstructed structures (central arrow) and hallucinations due to the high noise level (right arrow). SGM-based reconstructions are the mean of 5 sampled reconstructions.}
    \label{fig:2D_headline_comparison}
\end{figure*}

\subsection{2D reconstruction}
Reconstructions in 2D were performed using simulated data (Section \ref{sec:sim_data}). To calculate the 2D bias-variance assessment in Section \ref{sec:bias-variance} and 2D likelihood-variance assessment in Section \ref{sec:mlem_vs_pll}, for each of 10 random seeds, noisy sinogram data was generated according to the Poisson noise model. Then, reconstructions were performed for each independent noisy realization, with bias and standard deviation calculated according to Reader and Ellis\cite{reader_bootstrap-optimised_2020}.

Where unspecified, results from SGM-based reconstruction represent the mean of 5 samples with different random seeds (obtained from the same fixed noisy measured data), reconstructed with 100 generative timesteps. $\delta = 0.2$ was used by default, as well as 20 iterations of gradient ascent per generative step with PET-DDS. Normalized root mean square error (NRMSE) and structural similarity index measure (SSIM) were chosen to assess the global reconstruction performance.

\subsection{3D reconstruction}

Reconstructions in 3D were performed from real clinical research data (see Section \ref{sec:clinical_data}). To match clinical software outputs, our ParallelProj forward operator was used without a PSF. To calculate the 3D likelihood-variance assessment in Section \ref{sec:likelihood-variance}, prompts and randoms were sampled at 10\% of counts assuming independent Poisson statistics, with smoothed randoms and scatter sinograms re-estimated using Siemens' scanner software. To accommodate the computational demands of 3D reconstruction, we perform only 25 diffusion steps per reconstruction, compute only a single reconstructed sample instead of a sample mean, and also use $ \delta = 1.0 $.

All experiments were conducted on an NVIDIA GeForce RTX 3090 with 24 GB GPU memory.

\section{Results}

\subsection{2D reconstruction performance} \label{results-2D-overall}

Table \ref{tbl:2D_quant_results} shows the quantitative performance of each reconstruction method, assessed against 10 central 2D slices through each of 13 test datasets and averaged over 3 independent realizations of noisy sinogram data. Optimal hyperparameters for each method were established using a hyperparameter sweep to minimize NRMSE on the validation dataset.

\begin{table}[t]
	\caption[]{Quantitative results for reconstruction methods applied to 2D $[^{18}$F]DPA-714 simulated data, with hyperparameters for each method chosen to minimize NRMSE on a validation dataset. 95\% confidence intervals ($\pm$) calculated with respect to 3 different realizations of noisy measured data.}
	\label{tbl:2D_quant_results}
	\begin{center}
		\begin{sc}
            \input{webbe1.t}
		\end{sc}
	\end{center}
\end{table}

Fig. \ref{fig:2D_headline_comparison} presents representative reconstructed images from each method with these hyperparameters. As anticipated, our method and PET-DDS have similar performance quantitatively and qualitatively. We can say with confidence that without restrictions on time or hyperparameter searching, our method's reconstruction accuracy is at least on par with PET-DDS.

FBSEM-net-adv was the best-performing method in terms of both NRMSE and SSIM, which we attribute to the additional information incorporated in its training. In comparison, FBSEM-net over smooths reconstructions as a result of its simpler neural architecture.

The count level used is sufficiently low that several brain structures are not reconstructed by OSEM or MAP-EM. In these areas, the SGM-based reconstructions have more fine detail than other methods, but also more hallucinations; in contrast, FBSEM-net-adv has smoothed such areas. This difference in failure mode has contributed to the increased quantitative performance of FBSEM-net-adv relative to the SGM methods.

The indicative reconstruction times listed in Table \ref{tbl:2D_quant_results} allow us to conclude that the SGM-based methods are currently an order of magnitude slower than the conventional and supervised methods. For this set of optimal hyperparameters for PET-DDS, our proposed method was faster; this may not hold true in other settings. 

\begin{figure}
    \centering
    \includegraphics[width=\linewidth]{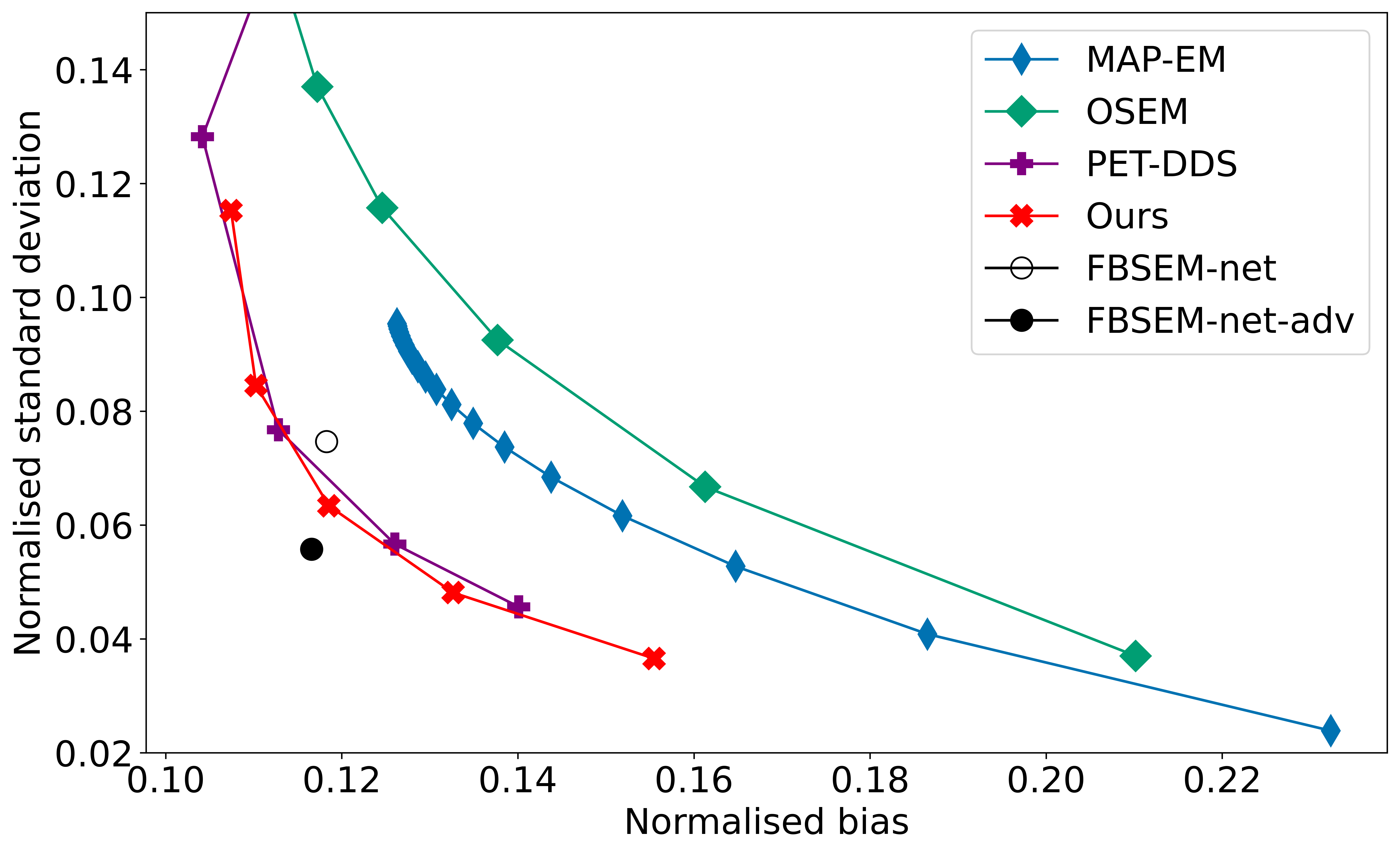}
    \caption{2D bias-variance assessment for each method \cite{reader_bootstrap-optimised_2020}. For OSEM and MAP-EM ($\beta = 3$), one subset was used and iteration number was varied from 5 to 100. For PET-DDS, $\lambda$ was varied from 0 to 2 (with $\lambda=0$ the rightmost point and $\lambda$ values increasing in increments of 0.5 for each of the data points displayed), whereas for our method $\nmlem$ was varied from 9 to 17 (increments of 2 shown, rightmost point corresponds to $\nmlem = 9$.}
    \label{fig:2D_bias_variance_chart}
\end{figure}

\subsection{2D bias-variance assessment} \label{sec:bias-variance}

Fig. \ref{fig:2D_bias_variance_chart} shows the results of a bias-variance assessment, performed on 10 central axial slices through the validation dataset. Where applicable, reconstruction hyperparameters were varied to show the effect of balancing the prior with the likelihood on the bias and variance properties of the reconstructions. This chart agrees with the previous quantitative results in Table \ref{tbl:2D_quant_results}. In particular, our method achieves a similar or superior bias to PET-DDS with optimal hyperparameters for all variance levels.

\subsection{2D hyperparameter stability} \label{sec:hyperparam_stability}

\begin{figure}
    \centering
    \includegraphics[width=\linewidth]{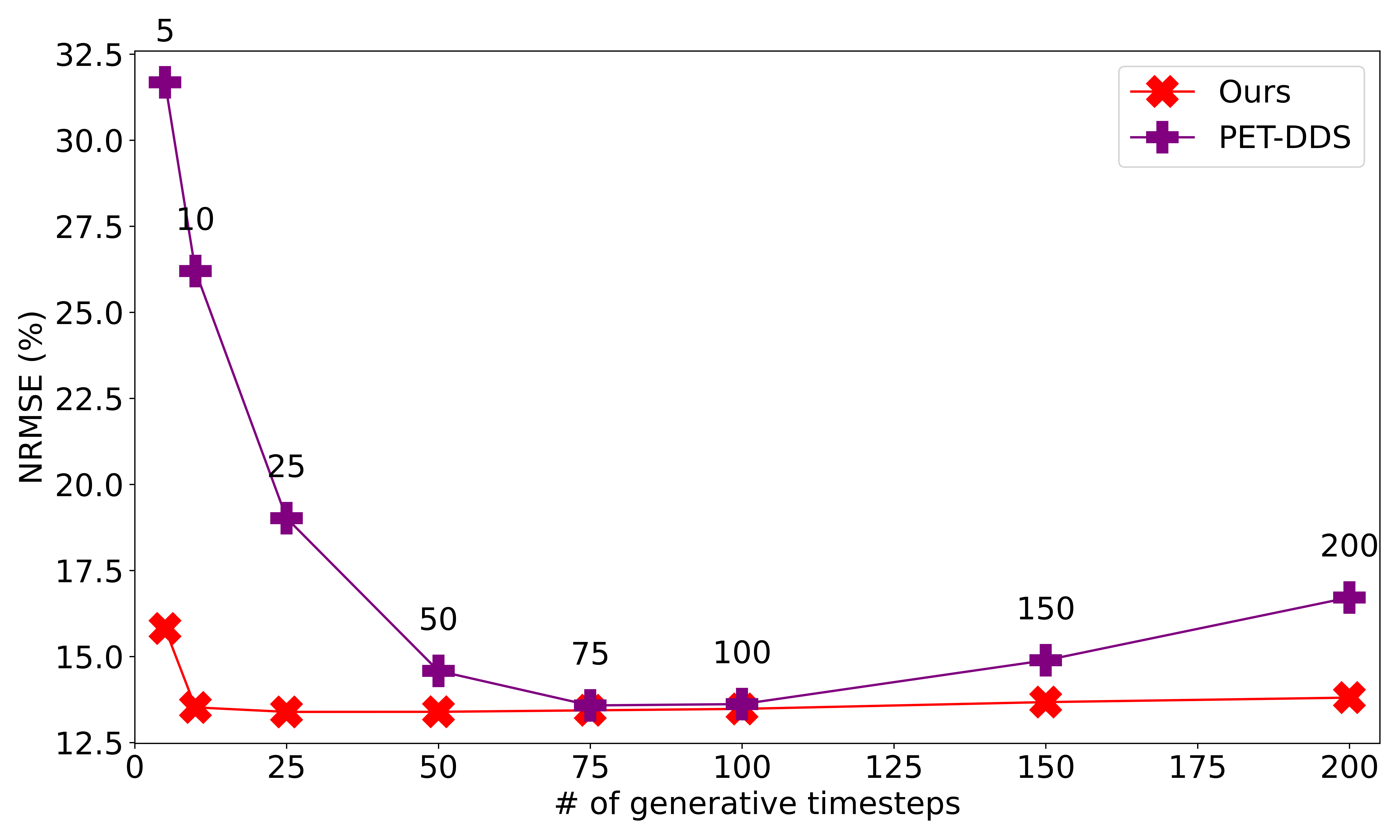}
    \caption{Reconstruction error (NRMSE) of 10 2D slices using optimal hyperparameters chosen for 100 generative timesteps at alternate numbers of generative timesteps for our method and PET-DDS.}
    \label{fig:dds_gen_steps}
\end{figure}

\begin{figure}[t!]
    \centering
    \includegraphics[width=\linewidth]{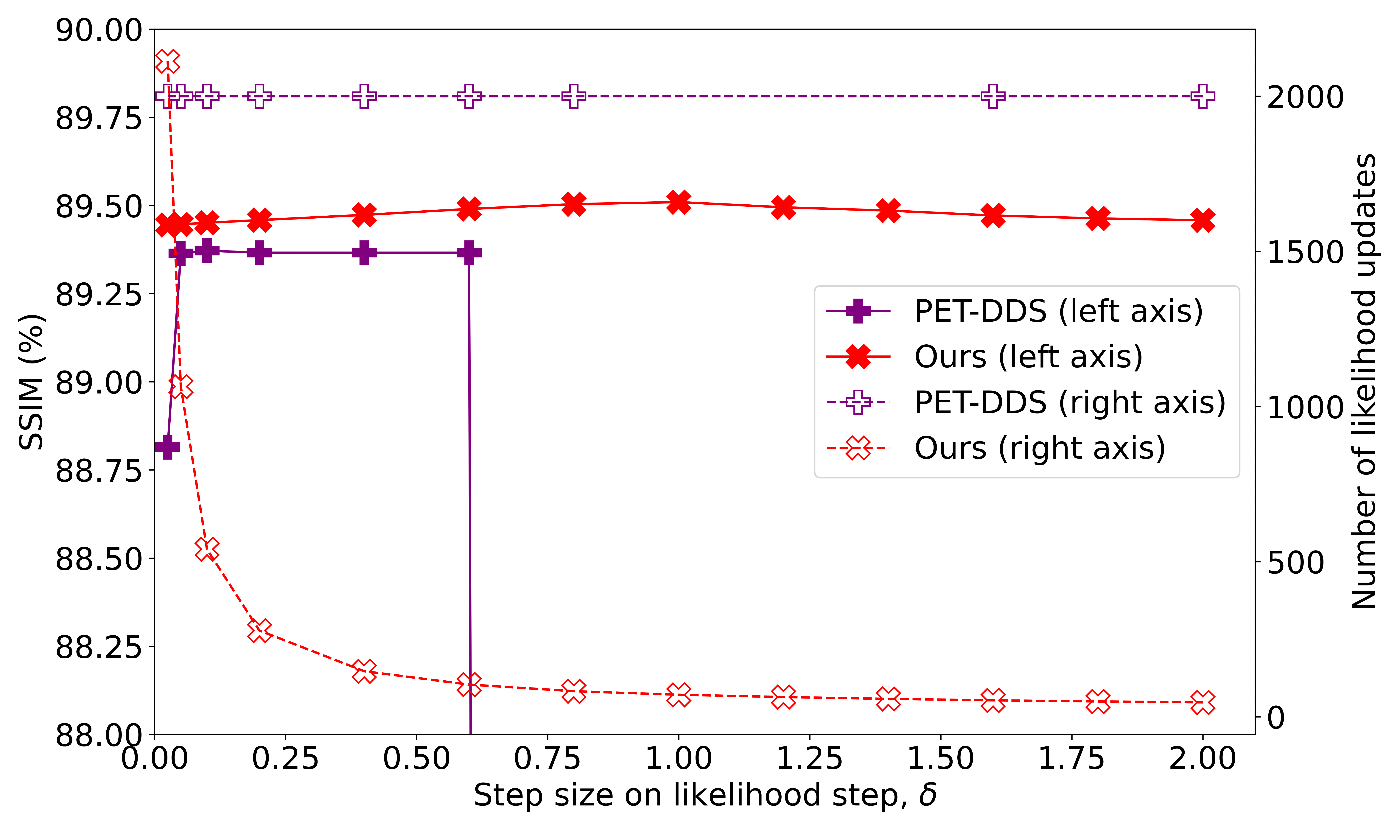}
    \caption{Reconstruction quality (SSIM) and number of likelihood updates for our method and PET-DDS using optimal hyperparameters chosen for 100 generative timesteps at alternate step sizes $\delta$, as evaluated on 10 2D slices.}
    \label{fig:varying_delta}
\end{figure}

To assess the stability of each SGM method, we considered varying the number of generative timesteps from 5 to 200 on both our method and PET-DDS, with the effect on reconstructions shown in Fig. \ref{fig:dds_gen_steps}. We also considered the effect of varying the step size $\delta$ of gradient ascent steps, shown in Fig. \ref{fig:varying_delta}. Whereas PET-DDS fails to converge for $\delta \ge 0.8$, our method is robust to at least $\delta \le 2.0$. Furthermore, our method with large $\delta$ uses fewer likelihood updates than generative steps, exhibiting remarkable efficiency, and demonstrating that our method is more efficiently exploiting the reduced need for gradient ascent steps with a strong diffusion prior.

These results show that our methodology is robust to both the number of generative timesteps chosen and the step size of gradient ascent employed. Therefore, for reasonable choices of generative timestep number and gradient ascent step size, our reconstruction error is solely a function of the target likelihood schedule.

\subsection{Sample path of likelihood-matched vs fixed updates}

\begin{figure*}
    \centering
    \begin{minipage}[t]{0.49\textwidth}
        \centering
    \includegraphics[width=\linewidth]{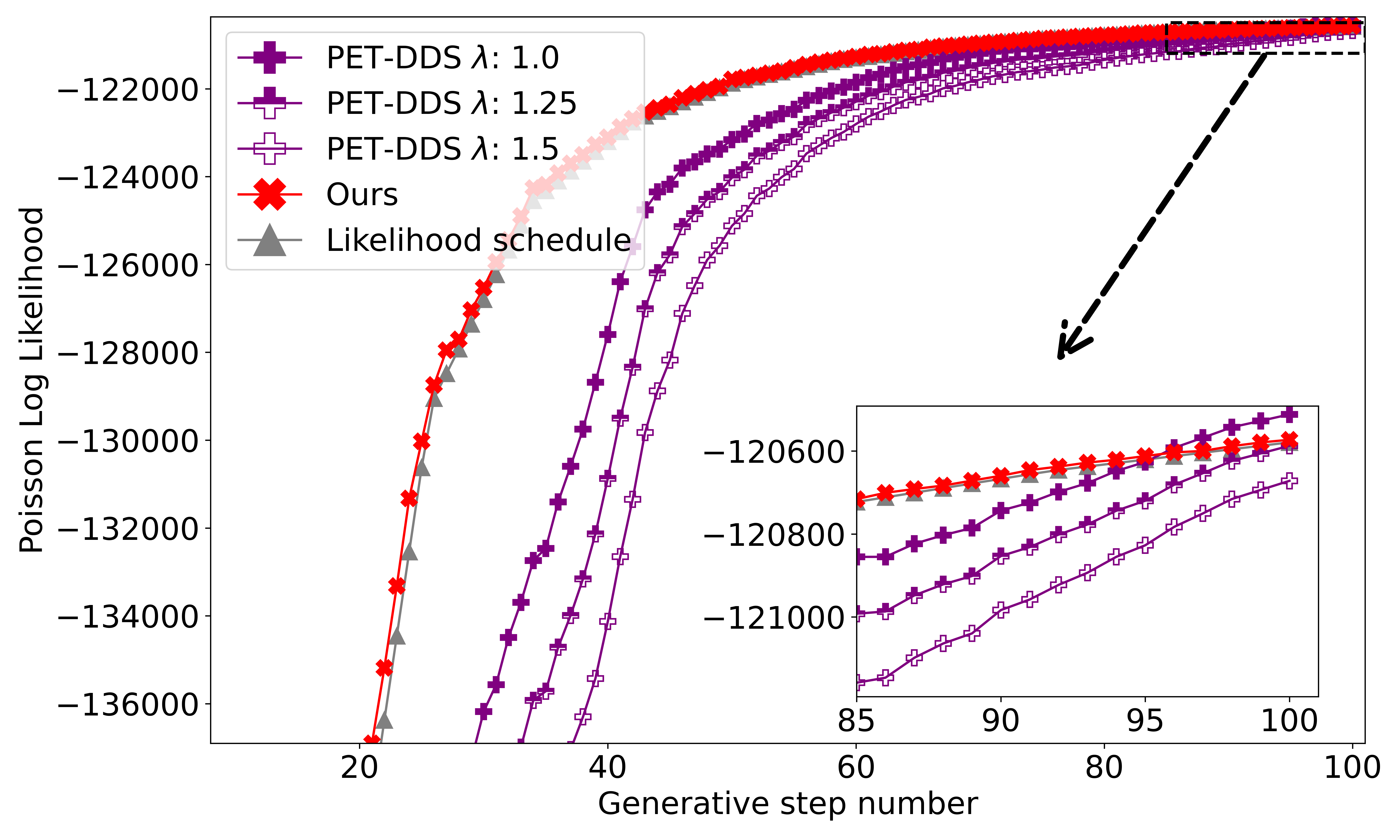}
        \caption{Example log-likelihood of measured data with respect to current Tweedie's estimate for a single reconstruction with our method and PET-DDS (with different values of $\lambda_{\text{MAP}}$).}
    \label{fig:likelihood_chart}
    \end{minipage}
    ~ 
    \begin{minipage}[t]{0.49\textwidth}
        \centering
        \includegraphics[width=\linewidth]{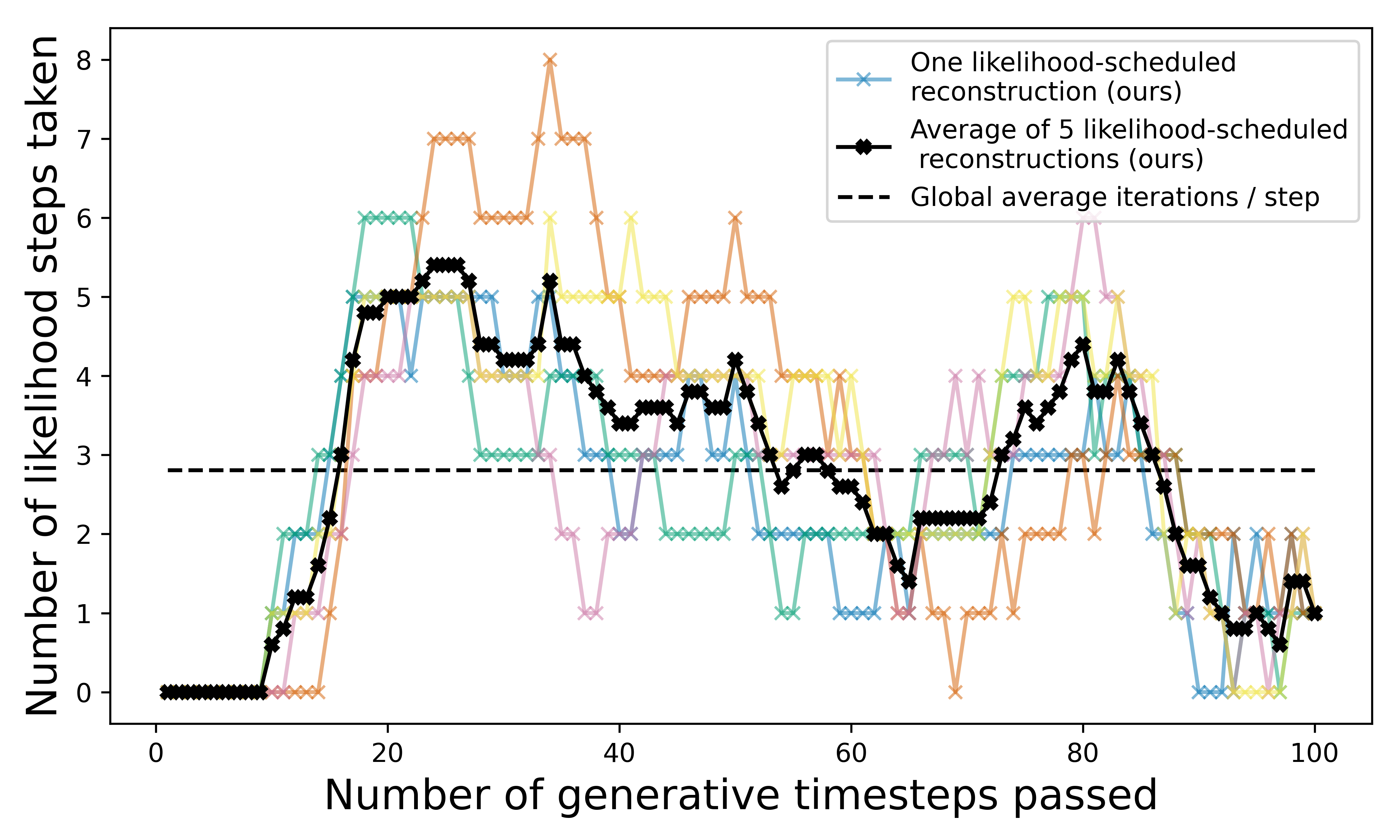}
     \caption{Number of likelihood steps taken per generative denoising step for a representative 2D PET reconstruction with likelihood-scheduled SGM-based PET reconstruction.}
     \label{fig:pls_steps}
    \end{minipage}
\end{figure*}

Fig. \ref{fig:likelihood_chart} compares the evolution of likelihood of Tweedie's estimate through the reverse diffusion process for our method and PET-DDS. We can see that the likelihood scheduling approach matches that of its likelihood schedule (and therefore the relevant MLEM estimate), whereas the PET-DDS reconstruction has no such guidance and with poor selection of $\lambda_{\text{MAP}}$ or $\delta$ is liable to overfit to noise or underfit to measurement data.

Fig. \ref{fig:pls_steps} shows the number of likelihood steps taken as a function of generative timestep. Whereas PET-DDS maintains a constant number of likelihood steps per generative timestep (reported by Singh \textit{et al.} from 4 to 20 \cite{singh_score-based_2024})), our method varies the number of likelihood steps to conform to the likelihood schedule. Relatively fewer steps at the start of the reverse diffusion process wastes less computation, as much information is lost when the re-noising step adds high-variance Gaussian noise to the Tweedie estimate. Fewer steps at the end of the reverse diffusion process reduces the chance of over-convergence to noisy measurement data.

In Section \ref{results-2D-overall}, our method used a mean of 201 likelihood updates per reconstructed sample (plus 14 for the likelihood scheduling) compared to 400-2000 fixed likelihood updates for PET-DDS (depending on the number of likelihood steps per generative timestep, reported from 4 to 20 \cite{singh_score-based_2024}). It is clear that there are potential efficiency gains to be made by dynamically varying the number of likelihood steps taken.

\begin{figure*}[th]
    \centering
    \includegraphics[width=\textwidth]{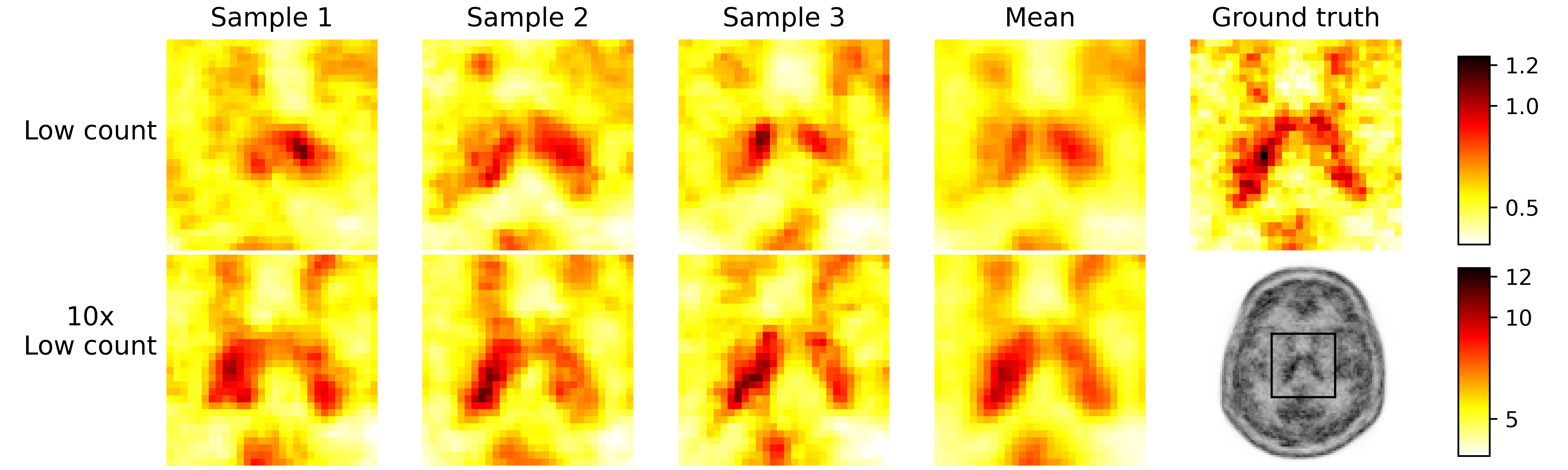}
    \caption{Example reconstruction samples from two different count levels, using different random seeds (but with fixed noisy data realizations for each count level).}
    \label{fig:2D_varied_recons}
\end{figure*}

\subsection{Effect on lesion recovery}
\begin{figure}
    \centering
    \includegraphics[width=\linewidth]{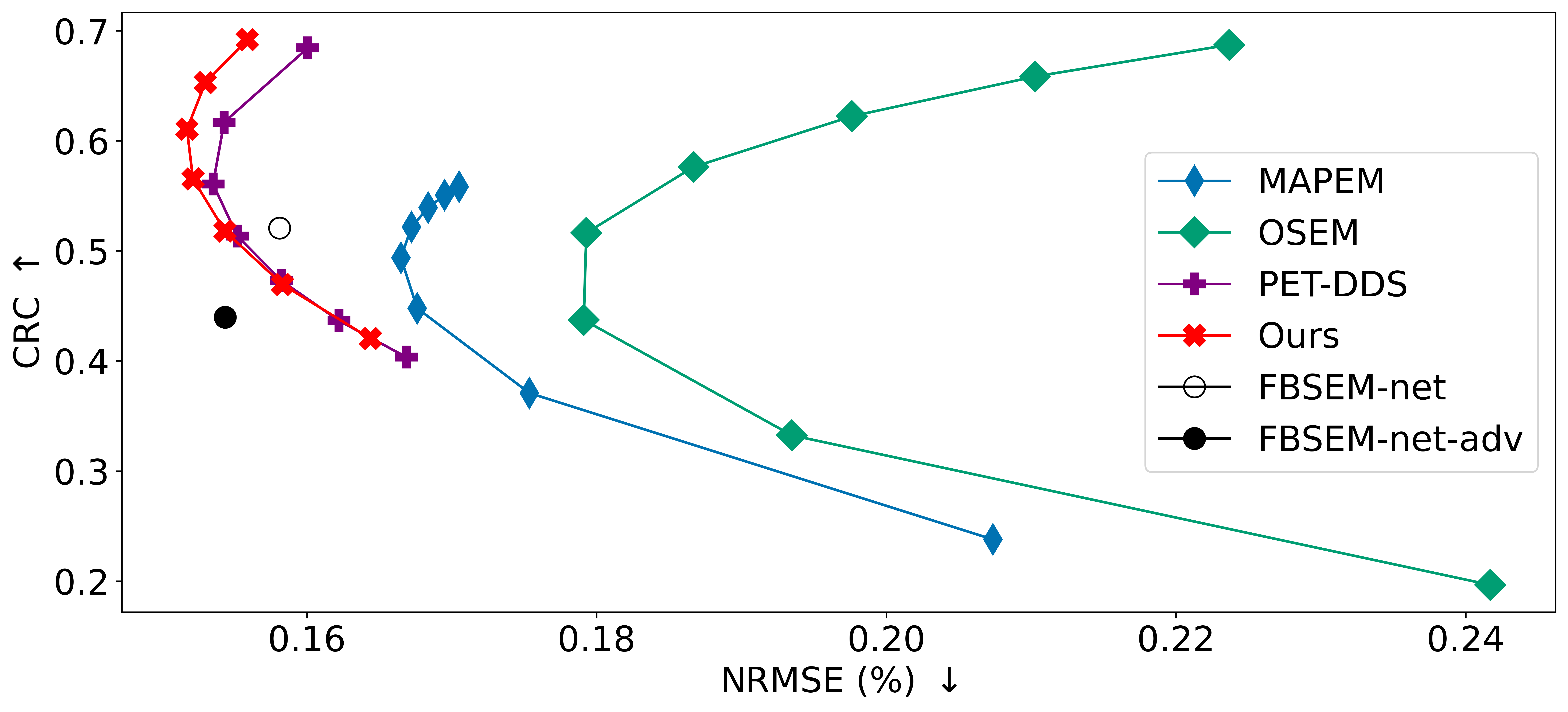}
    \caption{Evaluation of methods for a lesion recovery task, with an out-of-distribution hot lesion inserted into a test dataset. NRMSE is measured globally, while CRC is evaluated on the lesion itself, with a large area of white matter non-overlapping the lesion chosen as the background region. OSEM and MAP-EM ($\beta = 3$) have iteration number varied from 5 to 40 and 10 to 100 respectively. For PET-DDS, $\lambda$ was varied from 0.5 to 2.25, whereas for our method $N_{MLEM}$ was varied from 11 to 18. Results were averaged over 7 Poisson noise realizations. See Fig. \ref{fig:hot_spot_images} for corresponding images.}
    \label{fig:hot_spot_chart}
\end{figure}

\begin{figure*}
    \centering
    \includegraphics[width=\linewidth]{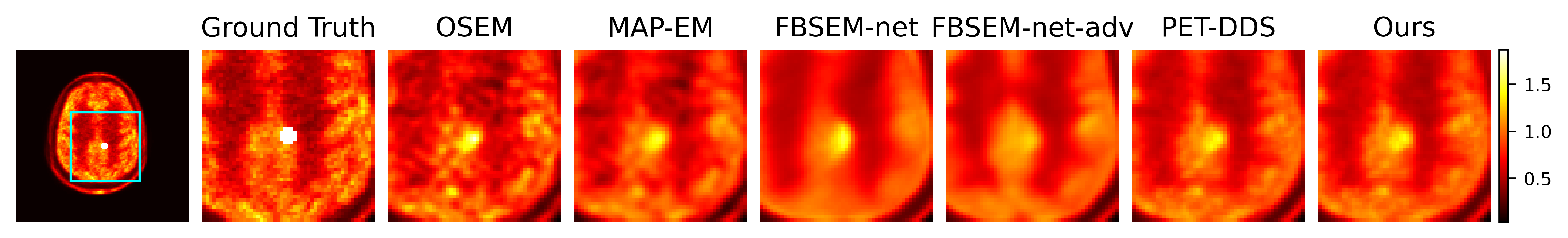}
    \caption{Visualization of each method's performance for resolving an out-of-distribution hot lesion inserted into a test dataset. Where hyperparameters were varied, the image for each method was chosen as the image with the best NRMSE for a CRC value of at least 0.55. See Fig. \ref{fig:hot_spot_chart} for each method's corresponding quantitative performance.}
    \label{fig:hot_spot_images}
\end{figure*}

A major strength of unsupervised diffusion model approaches relative to supervised approaches such as FBSEM-net is their ability to resolve out-of-distribution features such as lesions \cite{singh_score-based_2024}. To test this, we inserted a hot lesion into a 2D test dataset on the boundary between gray and white matter. We then performed reconstruction with this out-of-distribution dataset to investigate the lesion recovery performance of our approach. Figures \ref{fig:hot_spot_chart} and \ref{fig:hot_spot_images} verify that our approach does not adversely affect lesion recovery and that our method matches or outperforms the improvements to the contrast recovery coefficient (CRC) that are claimed by PET-DDS. FBSEM-net-adv (FBSEM-net with an advanced neural network) performs notably poorly at this task, highlighting that unsupervised approaches display greater flexibility to reconstructing out-of-distribution datasets (and thereby achieve superior lesion recovery) compared to supervised approaches such as FBSEM-net.

\subsection{2D reconstruction uncertainty}

For a fixed Poisson noise realization, we sample reconstructions based on the score-based prior. Fig. \ref{fig:2D_varied_recons} shows varied reconstructions from simulated data at two low count levels. As counts increase, less shape variation, closer resemblance to the ground truth, and fewer artifacts are observed. At both count levels, the mean image appears smoother and better matches the ground truth than the samples. The impact of dose level and generative timesteps on SGM performance will be explored in future work.

\subsection{Comparison to MLEM for varying iteration}\label{sec:mlem_vs_pll}

\begin{figure*}
    \centering
    \includegraphics[width=\linewidth]{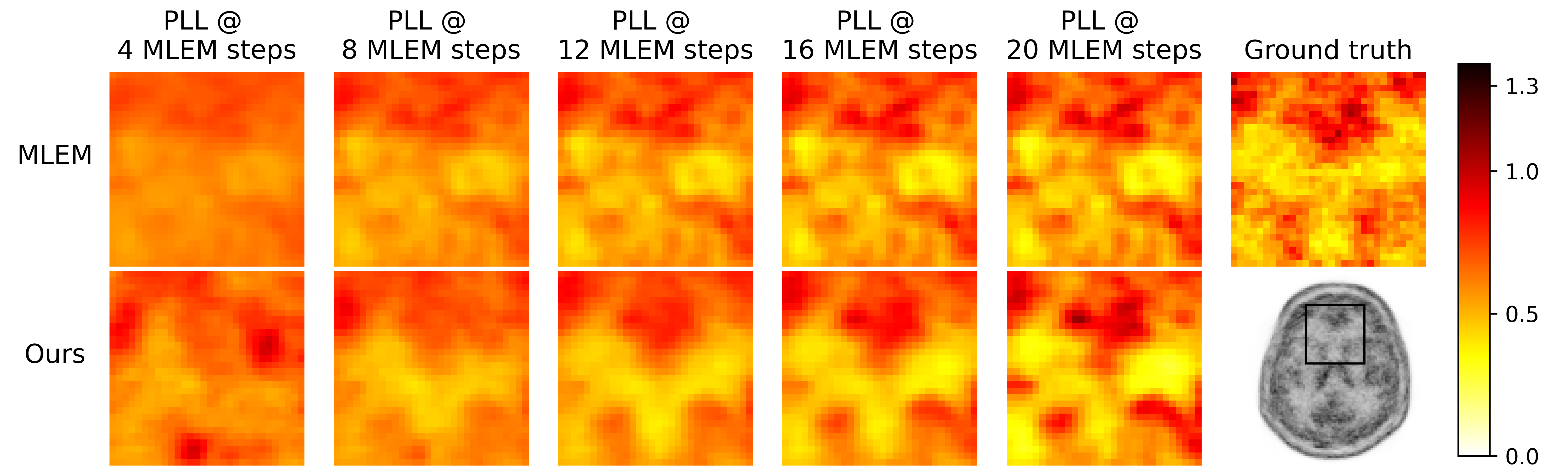}
    \caption{Example reconstructions with equivalent PLL using our likelihood-scheduling SGM-based method (mean of 5 samples) and MLEM, from 2D simulated data with 2.7 $\times 10^5$ counts.}
    \label{fig:2D_fixed_pll}
\end{figure*}

In Fig. \ref{fig:2D_fixed_pll}, we directly compared reconstructions with our methodology and the MLEM images used to derive their likelihood schedules. Our SGM-based methodology delivers noise reduction and better structure preservation than MLEM. As likelihood increases, both reconstructions become noisy and the noise pattern in our reconstructions closely matches the noise in the MLEM reconstruction.

\begin{figure*}[h!]
    \centering
    \includegraphics[width=\textwidth]{webbe13.png}
    \caption{Example reconstructions for real $[^{18}$F]DPA-714 data in 3D. All columns except ``Clinical" used the ParallelProj projector \cite{schramm_parallelprojopen-source_2024} \textit{without} PSF; ``Clinical" reconstructions used Siemens' proprietary tools. Reconstructions from 10\% count data match the PLL corresponding to 21 iterations of MLEM.}
    \label{fig:3D_real_data_images}
\end{figure*}

\begin{figure}
    \centering
    \includegraphics[width=\linewidth]{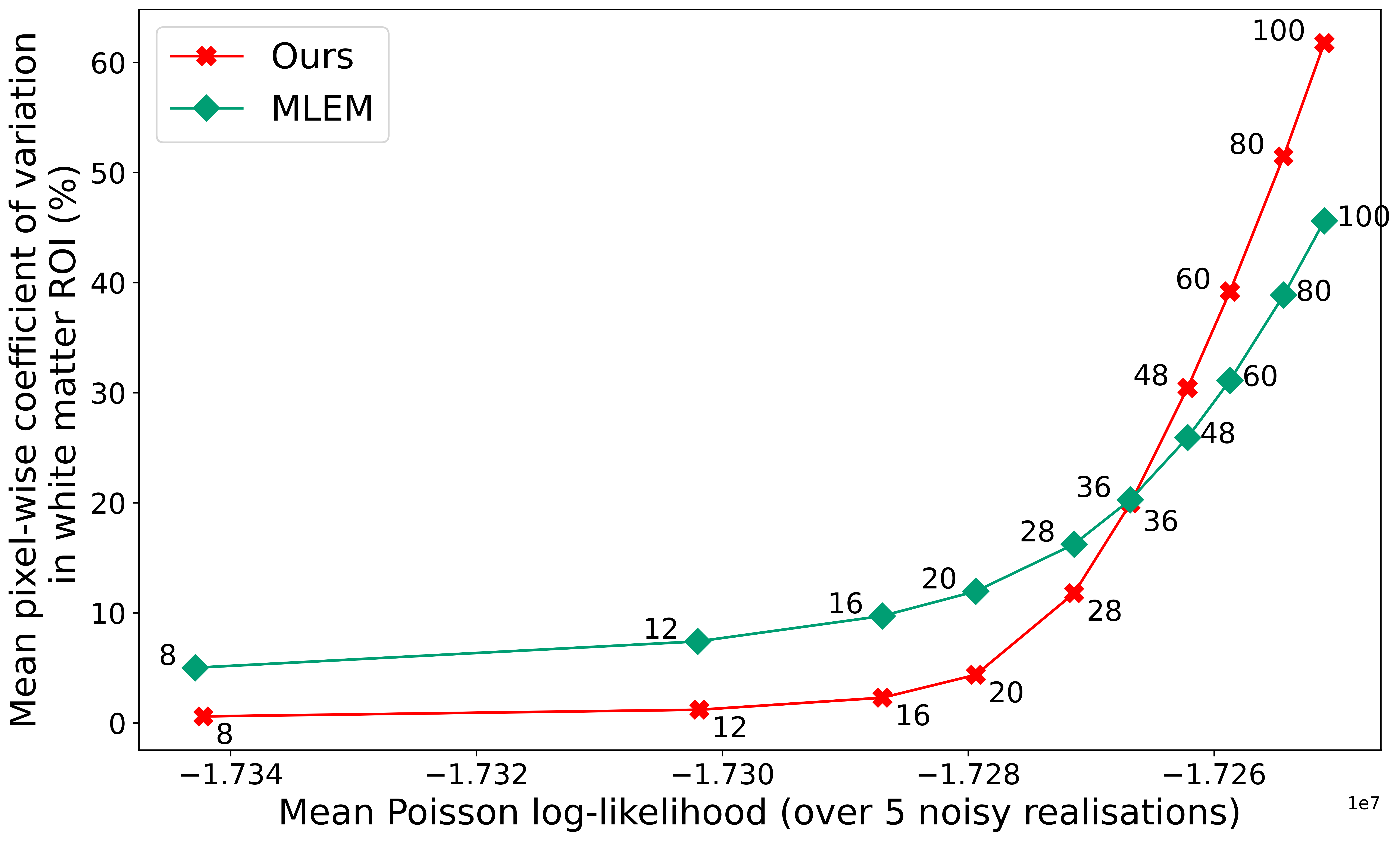}
    \caption{Coefficient of variation (CoV) against PLL for reconstructions from 10\% counts of real $[^{18}$F]DPA-714 3D measured data. CoV is measured as the mean pixel-wise coefficient of variation across reconstructions from 5 realizations of noisy measured data, as measured in a region of white matter (selected as a large visually uniform rectangle from the ground truth).}
    \label{fig:3D_real_data_quant}
\end{figure}

\subsection{Real 3D data}

\subsubsection{Qualitative results}

In Fig. \ref{fig:3D_real_data_images} we show reconstructions that achieve equivalent likelihood (to 21 iterations of MLEM) with different methods from real 10\% count data. We note that our introduction of perpendicular SGMs resolves the slice inconsistency (seen as alternating intensity transverse slices) seen for PET-DDS's (with a single SGM) coronal and sagittal slices. Both of the SGM-based methods display lower noise than the MLEM reconstruction at this likelihood level. (It should be noted that PET-DDS can eliminate slice inconsistency at lower likelihood at the cost of a blurring effect from the axial RDP.)

\subsubsection{Quantitative results}\label{sec:likelihood-variance}

Our methodology allows for direct comparisons between MLEM reconstruction (the clinical standard) and SGM-based reconstruction at the same likelihood. We leveraged this capability to assess the trade-off between likelihood and pixel-level variations between reconstructions of independent noise realizations of 10\% count data, shown in Fig. \ref{fig:3D_real_data_quant}. Our findings with real data concord with those in simulations, namely that for low likelihood values our methodology delivers improved reconstructions with reduced noise (relative to MLEM). In the regime of reconstructions over-fitting to noise ($\nmlem > 36$), reconstructions from our methodology vary more than MLEM; this may be because there do not exist images with such high likelihood on the SGM's manifold of probable images. However, all images at this likelihood level are too noisy to be suited to clinical tasks.

\begin{table}[t]
	\caption[]{Real computing time in seconds for computing a single sample with selected methods in 3D. PET-DDS was evaluated with 4 gradient-based steps per diffusion iteration, while our approach was used with $\nmlem = 21$ (corresponding to images in Fig. \ref{fig:3D_real_data_images}). Timings include constructing the likelihood schedule with 21 steps of MLEM for our approach.}
	\label{tbl:timing}
	\begin{center}
		\begin{sc}
            \input{webbe2.t}
		\end{sc}
	\end{center}
\end{table}

\subsubsection{Timing} We report the real computing time for 3D reconstruction with our proposed method and PET-DDS (both with and without the use of perpendicular SGMs) in Table \ref{tbl:timing}. Our approach is slightly more time efficient than PET-DDS for low-dose PET reconstruction (on a fair comparison with the same perpendicular pre-trained SGMs), with the overhead introduced from SGM steps greater than in 2D. The times for our approach include the time taken to construct the likelihood schedule; as this only occurs once per dataset, computing more samples is even more efficient with our method.

\section{Discussion}

Our results demonstrate that our method has a significantly lower hyperparameter tuning burden than PET-DDS, with the option of just tuning a single hyperparameter $\nmlem$ to directly vary the balance between the prior and the likelihood. This replaces tuning strength of MAP regularization $\lambda_{\text{MAP}}$; number of MAP iterations per generative step $N_{\text{MAP}}$; strength of RDP regularization $\lambda_{\text{RDP}}$ and gradient ascent step size $\delta$. In particular, this work integrates SGM priors with the standard clinical heuristic and vendor recommendation for the number of MLEM iterations.

Hallucinations in reconstructions are a known concern with SGM-based reconstruction. While hallucinations are present in some of the example low-dose reconstructions shown, it should be noted that this is an expected result; count levels were set deliberately low to explore the setting where structures are not clearly discernible from the OSEM reconstructions (and therefore cannot be easily reconstructed by MAP-EM even when relying on edge preservation priors). While individual samples may vary, given enough samples we can obtain a lower variance mean estimate. Our likelihood scheduling approach also makes it easier to increase the level of consistency with a model-based reconstruction by increasing $\nmlem$ (and potentially spot hallucinations); it could also be integrated with recent approaches for reducing hallucinations on out-of-distribution data \cite{barbano_steerable_2025}.

This work is the first methodology to investigate possible reconstructions for a fixed likelihood, providing the posterior distribution of image reconstruction conditioned on both a likelihood value and noisy measured data. This development opens the possibility of assessing the uncertainty of a reconstruction for a fixed likelihood level equivalent to a standard clinical reconstructed image. Furthermore, hyperparameter tuning could be eliminated completely by integrating this work with bootstrapping approaches for estimating the optimal PLL value or MLEM iteration \cite{reader_bootstrap-optimised_2020}. 

Other methods of deriving a likelihood schedule may provide superior efficiency to the MLEM-based method proposed, as could altering the SGM's noise schedule. Furthermore, the step size hyperparameter $\delta$ could be fully eliminated with an adaptive step size strategy or replacing gradient ascent with a different model-based step.

Further work on 3D modeling with pre-trained SGMs could improve the image quality in 3D further, for example by integrating our likelihood scheduling approach with 2.5D training approaches, latent diffusion models or patch-based approaches \cite{hu_patch-based_2024}. Lastly, the methods presented may find use cases in other medical image modalities where model-based iterative steps are combined with pre-trained SGMs, such as magnetic resonance imaging or computed tomography \cite{webber_diffusion_2024}.

\section{Conclusion}

In summary, we have shown a novel method for PET reconstruction with pre-trained SGMs, with the advantages of a lower hyperparameter tuning burden than previous SGM-based methods and simpler comparison to clinical methods. We further showed the applicability of pre-trained SGMs to real 3D PET reconstruction, and reduced issues with slice inconsistency and blurring in 3D.

\bibliographystyle{ieeetr}
\bibliography{tmi_author_accepted}

\end{document}

%% file: webbe1.t.tex
\begin{tabularx}{\linewidth}{l|YYY}
    \toprule
    Method & NRMSE (\%) $\downarrow$ & SSIM (\%) $\uparrow$ & Time (s)\\
    \midrule
    OSEM & 17.55 $\pm$ 0.10 & 84.18 $\pm$ 0.09 & 0.1\\
    MAP-EM & 17.02 $\pm$ 0.12 & 85.36 $\pm$ 0.14 & 0.3\\
    \midrule
    FBSEM-net & 15.05 $\pm$ 0.15 & 86.85 $\pm$ 0.10 & 0.2\\
    FBSEM-net-adv & 14.10 $\pm$ 0.03 & 88.45 $\pm$ 0.05 & 0.7\\
    \midrule
    PET-DDS & 14.82 $\pm$ 0.16 & 88.04 $\pm$ 0.12 &35\\
    Ours & 14.69 $\pm$ 0.13 & 88.02 $\pm$ 0.09 & 13\\
    \bottomrule
\end{tabularx}

%% file: webbe2.t.tex
\begin{tabularx}{\linewidth}{l|c|YY}
    \toprule
    \multirow{2}{*}{Method} & \multicolumn{2}{c}{Time (s)} \\
    \cmidrule(lr){2-3}
    & 25 SGM steps & 100 SGM steps \\
    \midrule
    MLEM & 33 & 33\\
    PET-DDS with single SGM & 116 & 398\\
    PET-DDS with perp. SGMs & 137 & 488\\
    Ours ($\delta = 1.0$) & 145 & 385\\
    Ours ($\delta = 2.0$) & 126 & 369\\
    Ours ($\delta = 4.0$) & 116 & 368\\
    \bottomrule
\end{tabularx}

%% file: tmi_author_accepted.bbl
\begin{thebibliography}{10}

\bibitem{bailey_positron_2005}
D.~L. Bailey, ed., {\em Positron emission tomography: basic sciences}.
\newblock New York: Springer, 2005.

\bibitem{boellaard_standards_2009}
R.~Boellaard, ``Standards for {PET} image acquisition and quantitative data analysis,'' {\em J Nucl Med}, vol.~50 Suppl 1, pp.~11S--20S, May 2009.

\bibitem{reader_ai_2023}
A.~J. Reader and B.~Pan, ``{AI} for {PET} image reconstruction,'' {\em British Journal of Radiology}, vol.~96, p.~20230292, Oct. 2023.

\bibitem{reader_deep_2021}
A.~J. Reader, G.~Corda, A.~Mehranian, C.~d. Costa-Luis, S.~Ellis, and J.~A. Schnabel, ``Deep {Learning} for {PET} {Image} {Reconstruction},'' {\em IEEE Transactions on Radiation and Plasma Medical Sciences}, vol.~5, pp.~1--25, Jan. 2021.

\bibitem{mehranian_model-based_2020}
A.~Mehranian and A.~J. Reader, ``Model-{Based} {Deep} {Learning} {PET} {Image} {Reconstruction} {Using} {Forward}-{Backward} {Splitting} {Expectation}-{Maximization},'' {\em IEEE Trans Radiat Plasma Med Sci}, vol.~5, pp.~54--64, June 2020.

\bibitem{guazzo_learned_2021}
A.~Guazzo and M.~Colarieti-Tosti, ``Learned {Primal} {Dual} {Reconstruction} for {PET},'' {\em J Imaging}, vol.~7, p.~248, Nov. 2021.

\bibitem{zhu_image_2018}
B.~Zhu, J.~Z. Liu, S.~F. Cauley, B.~R. Rosen, and M.~S. Rosen, ``Image reconstruction by domain-transform manifold learning,'' {\em Nature}, vol.~555, pp.~487--492, Mar. 2018.
\newblock Publisher: Nature Publishing Group.

\bibitem{haggstrom_deeppet_2019}
I.~Häggström, C.~R. Schmidtlein, G.~Campanella, and T.~J. Fuchs, ``{DeepPET}: {A} deep encoder–decoder network for directly solving the {PET} image reconstruction inverse problem,'' {\em Medical Image Analysis}, vol.~54, pp.~253--262, May 2019.

\bibitem{webber_diffusion_2024}
G.~Webber and A.~J. Reader, ``Diffusion {Models} for {Medical} {Image} {Reconstruction},'' {\em BJR{\textbar}Artificial Intelligence}, p.~ubae013, Aug. 2024.

\bibitem{chung_score-based_2022}
H.~Chung and J.~C. Ye, ``Score-based diffusion models for accelerated {MRI},'' {\em Medical Image Analysis}, vol.~80, p.~102479, Aug. 2022.

\bibitem{chung_solving_2023}
H.~Chung, D.~Ryu, M.~T. Mccann, M.~L. Klasky, and J.~C. Ye, ``Solving {3D} {Inverse} {Problems} {Using} {Pre}-{Trained} {2D} {Diffusion} {Models},'' in {\em 2023 {IEEE}/{CVF} {Conference} on {Computer} {Vision} and {Pattern} {Recognition} ({CVPR})}, (Vancouver, BC, Canada), pp.~22542--22551, IEEE, June 2023.

\bibitem{singh_score-based_2024}
I.~R. Singh {\em et~al.}, ``Score-{Based} {Generative} {Models} for {PET} {Image} {Reconstruction},'' {\em Melba}, vol.~2, pp.~547--585, Jan. 2024.

\bibitem{hu_unsupervised_2024}
R.~Hu {\em et~al.}, ``Unsupervised low-dose {PET} image reconstruction based on pre-trained denoising diffusion probabilistic prior,'' {\em Journal of Nuclear Medicine}, vol.~65, pp.~241109--241109, June 2024.

\bibitem{webber_generative-model-based_2024}
G.~Webber, Y.~Mizuno, O.~D. Howes, A.~Hammers, A.~P. King, and A.~J. Reader, ``Generative-{Model}-{Based} {Fully} {3D} {PET} {Image} {Reconstruction} by {Conditional} {Diffusion} {Sampling},'' in {\em 2024 {IEEE} {Nuclear} {Science} {Symposium} ({NSS}), {Medical} {Imaging} {Conference} ({MIC}) and {Room} {Temperature} {Semiconductor} {Detector} {Conference} ({RTSD})}, pp.~1--2, Oct. 2024.
\newblock ISSN: 2577-0829.

\bibitem{hou_fast_2024}
R.~Hou, F.~Li, and T.~Zeng, ``Fast and {Reliable} {Score}-{Based} {Generative} {Model} for {Parallel} {MRI},'' {\em IEEE Trans. Neural Netw. Learning Syst.}, pp.~1--14, 2024.

\bibitem{lee_improving_2023}
S.~Lee, H.~Chung, M.~Park, J.~Park, W.~Ryu, and J.~Ye, ``Improving {3D} {Imaging} with {Pre}-{Trained} {Perpendicular} {2D} {Diffusion} {Models},'' in {\em 2023 {IEEE}/{CVF} {International} {Conference} on {Computer} {Vision} ({ICCV})}, (Los Alamitos, CA, USA), pp.~10676--10686, IEEE Computer Society, Oct. 2023.

\bibitem{shepp_maximum_1982}
L.~A. Shepp and Y.~Vardi, ``Maximum {Likelihood} {Reconstruction} for {Emission} {Tomography},'' {\em IEEE Transactions on Medical Imaging}, vol.~1, pp.~113--122, Oct. 1982.

\bibitem{levitan_maximum_1987}
E.~Levitan and G.~T. Herman, ``A {Maximum} a {Posteriori} {Probability} {Expectation} {Maximization} {Algorithm} for {Image} {Reconstruction} in {Emission} {Tomography},'' {\em IEEE Transactions on Medical Imaging}, vol.~6, pp.~185--192, Sept. 1987.

\bibitem{hudson_accelerated_1994}
H.~M. Hudson and R.~S. Larkin, ``Accelerated image reconstruction using ordered subsets of projection data,'' {\em IEEE Trans Med Imaging}, vol.~13, no.~4, pp.~601--609, 1994.

\bibitem{de_pierro_fast_2001}
A.~De~Pierro and M.~Yamagishi, ``Fast {EM}-like methods for maximum "a posteriori" estimates in emission tomography,'' {\em IEEE Transactions on Medical Imaging}, vol.~20, pp.~280--288, Apr. 2001.

\bibitem{song_generative_2019}
Y.~Song and S.~Ermon, ``Generative {Modeling} by {Estimating} {Gradients} of the {Data} {Distribution},'' in {\em Advances in {Neural} {Information} {Processing} {Systems}}, vol.~32, Curran Associates, Inc., 2019.

\bibitem{sohl-dickstein_deep_2015}
J.~Sohl-Dickstein, E.~A. Weiss, N.~Maheswaranathan, and S.~Ganguli, ``Deep unsupervised learning using nonequilibrium thermodynamics,'' in {\em Proceedings of the 32nd {International} {Conference} on {International} {Conference} on {Machine} {Learning} - {Volume} 37}, {ICML}'15, (Lille, France), pp.~2256--2265, JMLR.org, July 2015.

\bibitem{ho_denoising_2020}
J.~Ho, A.~Jain, and P.~Abbeel, ``Denoising {Diffusion} {Probabilistic} {Models},'' in {\em Advances in {Neural} {Information} {Processing} {Systems}} (H.~Larochelle, M.~Ranzato, R.~Hadsell, M.~F. Balcan, and H.~Lin, eds.), vol.~33, pp.~6840--6851, Curran Associates, Inc., 2020.

\bibitem{song_score-based_2020}
Y.~Song, J.~Sohl-Dickstein, D.~P. Kingma, A.~Kumar, S.~Ermon, and B.~Poole, ``Score-{Based} {Generative} {Modeling} through {Stochastic} {Differential} {Equations},'' Oct. 2020.

\bibitem{anderson_reverse-time_1982}
B.~D. Anderson, ``Reverse-time diffusion equation models,'' {\em Stochastic Processes and their Applications}, vol.~12, pp.~313--326, May 1982.

\bibitem{vincent_connection_2011}
P.~Vincent, ``A {Connection} {Between} {Score} {Matching} and {Denoising} {Autoencoders},'' {\em Neural Computation}, vol.~23, pp.~1661--1674, July 2011.

\bibitem{efron_tweedies_2011}
B.~Efron, ``Tweedie's {Formula} and {Selection} {Bias},'' {\em Journal of the American Statistical Association}, vol.~106, no.~496, pp.~1602--1614, 2011.

\bibitem{chung_decomposed_2023}
H.~Chung, S.~Lee, and J.~C. Ye, ``Decomposed {Diffusion} {Sampler} for {Accelerating} {Large}-{Scale} {Inverse} {Problems},'' Oct. 2023.

\bibitem{zhu_denoising_2023}
Y.~Zhu {\em et~al.}, ``Denoising {Diffusion} {Models} for {Plug}-and-{Play} {Image} {Restoration},'' in {\em 2023 {IEEE}/{CVF} {Conference} on {Computer} {Vision} and {Pattern} {Recognition} {Workshops} ({CVPRW})}, pp.~1219--1229, June 2023.
\newblock ISSN: 2160-7516.

\bibitem{xie_joint_2024}
T.~Xie {\em et~al.}, ``Joint diffusion: mutual consistency-driven diffusion model for {PET}-{MRI} co-reconstruction,'' {\em Phys Med Biol}, vol.~69, July 2024.

\bibitem{wang_penalized_2012}
G.~Wang and J.~Qi, ``Penalized {Likelihood} {PET} {Image} {Reconstruction} using {Patch}-based {Edge}-preserving {Regularization},'' {\em IEEE Trans Med Imaging}, vol.~31, pp.~2194--2204, Dec. 2012.

\bibitem{de_pierro_relation_1993}
A.~R. De~Pierro, ``On the relation between the {ISRA} and the {EM} algorithm for positron emission tomography,'' {\em IEEE Trans Med Imaging}, vol.~12, no.~2, pp.~328--333, 1993.

\bibitem{schramm_parallelprojopen-source_2024}
G.~Schramm and K.~Thielemans, ``{PARALLELPROJ}—an open-source framework for fast calculation of projections in tomography,'' {\em Front. Nucl. Med.}, vol.~3, Jan. 2024.
\newblock Publisher: Frontiers.

\bibitem{belzunce_assessment_2017}
M.~A. Belzunce and A.~J. Reader, ``Assessment of the impact of modeling axial compression on {PET} image reconstruction,'' {\em Med Phys}, vol.~44, pp.~5172--5186, Oct. 2017.

\bibitem{muratib_dissection_2021}
F.~Muratib, Y.~Mizuno, I.~C. Figueiredo, O.~Howes, and T.~R. Marques, ``Dissection of neuroinflammation in schizophrenia,'' {\em BJPsych Open}, vol.~7, pp.~S274--S275, June 2021.

\bibitem{reader_bootstrap-optimised_2020}
A.~J. Reader and S.~Ellis, ``Bootstrap-{Optimised} {Regularised} {Image} {Reconstruction} for {Emission} {Tomography},'' {\em IEEE Transactions on Medical Imaging}, vol.~39, pp.~2163--2175, June 2020.

\bibitem{barbano_steerable_2025}
R.~Barbano, A.~Denker, H.~Chung, T.~H. Roh, S.~Arridge, P.~Maass, B.~Jin, and J.~C. Ye, ``Steerable {Conditional} {Diffusion} for {Out}-of-{Distribution} {Adaptation} in {Medical} {Image} {Reconstruction},'' {\em IEEE Transactions on Medical Imaging}, pp.~1--1, 2025.

\bibitem{hu_patch-based_2024}
J.~Hu, B.~Song, J.~A. Fessler, and L.~Shen, ``Patch-{Based} {Diffusion} {Models} {Beat} {Whole}-{Image} {Models} for {Mismatched} {Distribution} {Inverse} {Problems},'' Oct. 2024.
\newblock arXiv:2410.11730 [cs].

\end{thebibliography}
